\documentclass[preprint,12p]{elsarticle}

\usepackage{amsmath,amssymb,amsfonts,amssymb,stmaryrd}
\usepackage{bm}
\usepackage[utf8x]{inputenc}
\usepackage{tensor}
\usepackage{braket}
\usepackage{xcolor}
\usepackage{physics}
\usepackage{dsfont}
\usepackage{listings}
\usepackage{hyperref}

\newcommand{\lgr}{\mathcal{L}}

\newcommand{\D}{\textnormal{d}}
\newcommand{\Mpl}{M_{\rm{Pl}}}
\newcommand{\tMpl}{\tilde{M}_{\rm{Pl}}}
\newcommand{\code}[1]{\texttt{#1}}
\biboptions{sort&compress}

\definecolor{lightgrey}{gray}{0.9}
\def\btab#1\etab{\begin{tabular}{p{45mm}p{65mm}}#1\end{tabular}}
\def\btabx#1\etabx{\begin{tabular}{p{60mm}p{50mm}}#1\end{tabular}}
\def\btaby#1\etaby{\begin{tabular}{p{15mm}p{95mm}}#1\end{tabular}}
\def\bcen{\begin{center}}
\def\ecen{\end{center}}
\def\bgfb#1\egfb{\bcen\fcolorbox{black}{lightgrey}{\parbox{118mm}{\btab#1\etab}}\ecen}
\def\bgfbx#1\egfbx{\bcen\fcolorbox{black}{lightgrey}{\parbox{118mm}{\btabx#1\etabx}}\ecen}
\def\bgfbalign#1\egfbalign{\bcen\fcolorbox{black}{lightgrey}{\parbox{118mm}{\btaby#1\etaby}}\ecen}
\newcommand{\edit}[1]{#1}
\usepackage[T1]{fontenc}
\usepackage{lmodern}

\usepackage{mmacells}

\mmaDefineMathReplacement[≤]{<=}{\leq}
\mmaDefineMathReplacement[≥]{>=}{\geq}
\mmaDefineMathReplacement[≠]{!=}{\neq}
\mmaDefineMathReplacement[→]{->}{\to}[2]
\mmaDefineMathReplacement[⧴]{:>}{:\hspace{-.2em}\to}[2]
\mmaDefineMathReplacement{∉}{\notin}
\mmaDefineMathReplacement{∞}{\infty}
\mmaDefineMathReplacement{��}{\mathbbm{d}}

\mmaSet{
  morefv={gobble=2},
  linklocaluri=mma/symbol/definition:#1,
  morecellgraphics={yoffset=1.9ex}
}

\begin{document}
\begin{frontmatter}

\title{\vspace*{-30mm} \includegraphics[scale=0.1]{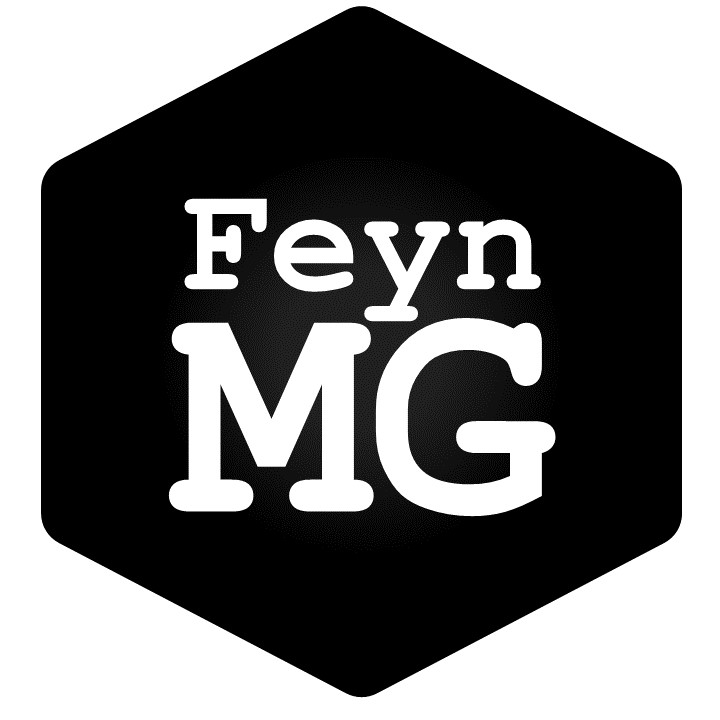}\\[6pt]
        \code{FeynMG}:~a \code{FeynRules} extension for scalar-tensor theories of gravity}

\author[a,b]{Sergio Sevillano Mu\~noz\corref{author}}
\author[a]{Edmund J. Copeland}
\author[c]{\\ Peter Millington}
\author[b]{Michael Spannowsky}

\cortext[author] {Corresponding author\\\textit{E-mail addresses:}\\ sergio.sevillano-munoz@durham.ac.uk (Sergio Sevillano Mu\~noz),\\
edmund.copeland@nottingham.ac.uk (Edmund J.~Copeland),\\peter.millington@manchester.ac.uk (Peter Millington),\\ michael.spannowsky@durham.ac.uk (Michael Spannowsky)\\
\\
\noindent This is an author-prepared post-print of \href{https://doi.org/10.1016/j.cpc.2023.109035}{Comput. Phys. Commun. 296 (2024) 109035}, published by Elsevier under the terms of the \href{https://creativecommons.org/licenses/by/4.0/}{CC BY 4.0} license.
}
\address[a]{School of Physics and Astronomy, University of Nottingham,\\
Nottingham, NG7 2RD, UK}
\address[b]{Institute for Particle Physics Phenomenology, Department of Physics, Durham
University,\\ Durham, DH1 3LE, UK}
\address[c]{Department of Physics and Astronomy, University of Manchester,\\
Manchester, M13 9PL, UK}
\vspace{-5mm}
\begin{abstract}
The ability to represent perturbative expansions of interacting quantum field theories in terms of simple diagrammatic rules has revolutionized calculations in particle physics (and elsewhere). Moreover, these rules are readily automated, a process that has catalysed the rise of symbolic algebra packages. However, in the case of extended theories of gravity, such as scalar-tensor theories, it is necessary to precondition the Lagrangian to apply this automation or, at the very least, to take advantage of existing software pipelines. We present a \code{Mathematica} code \texttt{FeynMG}, which works in conjunction with the well-known package \texttt{FeynRules}, to do just that:~\texttt{FeynMG} takes as inputs the \texttt{FeynRules} model file for a non-gravitational theory and a user-supplied gravitational Lagrangian. \texttt{FeynMG} provides functionality that inserts the minimal gravitational couplings of the degrees of freedom specified in the model file, determines the couplings of the additional tensor and scalar degrees of freedom (the metric and the scalar field from the gravitational sector), and preconditions the resulting Lagrangian so that it can be passed to \texttt{FeynRules}, either directly or by outputting an updated \texttt{FeynRules} model file. The Feynman rules can then be determined and output through \texttt{FeynRules}, using existing universal output formats and interfaces to other analysis packages.
\end{abstract}

\begin{keyword}
Feynman rules, symbolic algebra, scalar-tensor theories, high energy physics, FeynRules.
\end{keyword}
\end{frontmatter}


\noindent
{\bf PROGRAM SUMMARY} \\[-5pt]

\begin{small}
\noindent
{\em Program title:} FeynMG \\[5pt]
{\em CPC Library link to program files:} \href{https://doi.org/10.17632/bjpfmtt55k.1}{doi.org/10.17632/bjpfmtt55k.1}\\[5pt]
{\em Developer's repository link:} \href{https://gitlab.com/feynmg/FeynMG}{gitlab.com/feynmg/FeynMG} \\[5pt]
{\em Licensing provisions:} : MIT license (MIT)\\[5pt]
{\em Programming language:} \code{Mathematica 13.2}\\[5pt]
{\em Nature of the problem:} \\[2pt]
Determining the additional interactions that scalar-tensor theories of gravity induce between the fields of the Standard Model of particle physics and its extensions is a tedious and time-consuming task that is ripe for automation. \texttt{FeynMG} is a package that provides this automation. It can expand the spacetime metric around a flat background and perform the necessary field redefinitions for a given theory in order to provide a Lagrangian that is in a form that can be processed by \code{FeynRules}~\cite{feynrules}. The Feynman rules for the theory can then be determined using the existing functionality of \code{FeynRules} and output in formats that can be read into other particle-physics analysis packages.\\[5pt]

\noindent{\em Solution method:} \\[2pt]
(1) Load both \code{FeynRules} and \texttt{FeynMG} into \code{Mathematica}. (2)~Load a model file describing a given matter sector and Lagrangian. (3)~Use the \code{FeynMG} implementation to: (3a) append the gravitational sector and incorporate curvature-dependent objects, such as the metric, curvature scalars and tensors, and covariant derivatives; (3b)~effect a Weyl transformation or expand the spacetime metric up to second order in graviton interactions with the matter fields; and (3c)~digonalize mass and kinetic mixings, and canonically normalize dynamical fields. (4)~Pass the output directly to \code{FeynRules} or output a new model file from \code{FeynMG}.
\end{small}


\section{Introduction}

The increasing complementarity of high-precision data from cosmological observations and high energy physics experiments makes it necessary to consider non-minimal gravitational couplings or the impact of additional degrees of freedom that are coupled through the gravitational sector with strengths that need not be Planck-suppressed. Examples include scalar-tensor theories of gravity~\cite{FM}, such as the Brans--Dicke theory~\cite{Brans:1961sx} or, more generally, the Horndeski theories~\cite{Horndeski:1974wa, Kobayashi:2019hrl} (including beyond Horndeski~\cite{Traykova:2019oyx, Gleyzes:2013ooa} and DHOST~\cite{Langlois:2015cwa, Langlois:2018dxi} theories), in which the gravitational sector includes both the metric and an additional scalar degree of freedom. Other relevant examples include those in which the Higgs is non-minimally coupled to gravity, as is required in Higgs inflation~\cite{Wetterich:1987fm,Buchmuller:1988cj,Shaposhnikov:2008xb,Shaposhnikov:2008xi,Blas:2011ac,Henz:2013oxa,Karananas:2016grc,Ferreira:2016vsc} or so-called Higgs-Dilaton models~\cite{Rubio:2014wta,Bezrukov:2012hx,Garcia-Bellido:2012npk,Garcia-Bellido:2011kqb}. Indeed, such non-minimal couplings of the Higgs field to the scalar curvature are readily motivated by considering the renormalization-group evolution of the operators of the Standard Model of particle physics plus gravity~\cite{Herranen:2014cua, Markkanen:2018bfx,Steinwachs:2011zs}. Moreover, the ability to make Weyl rescalings of the metric and so-called disformal transformations~\cite{Bekenstein:1992pj,Zumalacarregui:2013pma,Teixeira:2019hil} allows us to make connections between scalar-tensor theories of gravity and gauge-singlet, scalar extensions of the Standard Model of particle physics, such as Higgs- or neutrino-portal theories~\cite{Silveira:1985rk, McDonald:1993ex, Burgess:2000yq, Davoudiasl:2004be, Schabinger:2005ei,Patt:2006fw,Falkowski:2009yz,GonzalezMacias:2015rxl}.

The challenge, however, is the proliferation of operators that non-minimal gravitational couplings provide, alongside degeneracies with operators that directly couple new degrees of freedom to the Standard Model. Dealing with this requires linearization of the extended gravitational sector, transformations of the metric, expansion around non-trivial vacuum configurations, the diagonalization of kinetic and mass mixings, and the truncation of infinite series of operators~\cite{Burrage:2018dvt,Copeland:2021qby}. This is usually done on a model-by-model basis, and it is a tedious and time-consuming process, which is ripe for automating, and doing so is the focus of this article.

We present a \code{Mathematica} package \code{FeynMG}, which is designed to work alongside the well-known \code{FeynRules} package~\cite{feynrules}. \texttt{FeynRules} is an extensive \code{Mathematica} package that enables the user to output the Feynman rules for a given Lagrangian in formats that can be read in by a range of high energy physics analysis software, including \code{CalcHep}/\code{CompHEP}~\cite{Belyaev:2012qa,Pukhov:1999gg}, \code{FeynArts}~\cite{Kublbeck:1990xc}, \code{FeynCalc}~\cite{Shtabovenko:2020gxv}, \code{FormCalc}~\cite{Hahn:2016ebn}, \code{MadGraph}~\cite{Alwall:2011uj}, \code{Sherpa}~\cite{Sherpa:2019gpd}, \code{Whizard/Omega}~\cite{Kilian:2007gr} and \code{ASperge}~\cite{Asperge}.

Symbolic algebra packages have also been developed to deal with the complex tensor algebra that arises in General Relativity. A recent example is \code{FeynGrav}~\cite{Latosh:2022ydd}, a package that introduces gravity in its canonical form (the Einstein--Hilbert action) to \code{FeynRules}. \code{xAct}~\cite{xAct} is perhaps the most well-known package, having already been followed by multiple compatible packages that allow the study of gravity in different cosmological scenarios. In particular, the package \code{xIST/COPPER}~\cite{xIST} extends \code{xAct} for general scalar-tensor theories, and it was used in Ref.~\cite{Lagos:2016wyv} to calculate the effect of modified gravity on cosmological perturbations. In this sense, \code{FeynMG} extends \code{FeynRules} as \code{xIST/COPPER} extends \code{xAct}. 

\texttt{FeynMG} is intended as a `preconditioner'. It takes as inputs a \texttt{FeynRules} model file and the Lagrangian of an extended gravitational sector. \texttt{FeynMG} then provides the functionality to implement the minimal gravitational couplings to the Lagrangian from the original model file and cast the complete theory in a form that can be further processed using the existing \texttt{FeynRules} package and its interfaces. However, we emphasise that \code{FeynMG} contains functionality that may be useful for theories that are being analysed independent of the couplings to gravitational sectors, and this will be highlighted throughout this article.

The remainder of this article is structured as follows. In Section~\ref{sec:method}, we describe the general form of the problem of coupling the Standard Model to extended gravitational sectors. We then present the package \code{FeynMG}, summarizing the implementation in Section~\ref{sec:implementation} and describing its usage in Section~\ref{sec:usage}. Finally, our conclusions are presented in Section~\ref{sec:conclusions}, and additional technical details are provided in the Appendices.

Throughout this work, while it is a convention that is uncommon in the gravitation and cosmology literature, we use the ``mostly minus'' metric signature convention $(+,-,-,-)$, in which timelike four-momenta $p^{\mu}$ have $p^2>0$, \edit{and the positive Clifford relation given by $\{\gamma^a,\gamma^b\}=2\eta^{ab}$}, since this is the convention commonly used by existing particle physics software packages. We use lower-case Greek characters for the Lorentz indices of the curved spacetime and lower-case Roman characters for the Lorentz indices of the flat, tangent space necessary for writing the Dirac Lagrangian in a generally covariant form. $D$ denotes gauge covariant derivatives, general (i.e., gravitational) and gauge covariant derivatives are denoted by $\nabla$, and an update to the general and gauge covariant derivative that is useful for scalar-tensor theories of Brans--Dicke type is represented by $\mathcal{D}$. We work in natural units, but \textit{do not} set Newton's gravitational constant to unity.
	

\section{Method}
\label{sec:method}

We begin by reviewing how a Minkowski quantum field theory is minimally coupled to gravity and how additional scalar fields that are non-minimally coupled to the scalar curvature of the gravity sector can give rise to new interactions in that quantum field theory.

For simplicity, we work with a toy model of QED plus a real scalar prototype of the Higgs sector. Generalizing to a complex scalar field that is charged under $U(1)$ would be a technical complication that does not add to the main points that we wish to illustrate below. The action of this model in Minkowski spacetime is given by
 \begin{align}
	     S_{\rm m}=&\int \D^4{x} \left[-\frac{1}{4}F^{\mu\nu}F_{\mu\nu}+\frac{1}{2}\partial_\mu \phi \partial^\mu \phi\right. \nonumber\\ 
	     &+i\bar{\psi}\gamma^{\mu}D_\mu\psi -y\bar{\psi}\phi\psi +\frac{1}{2}\mu^2 \phi^2 -\frac{\lambda}{4!}\phi^4 -\frac{3\mu^4}{2\lambda}\bigg],
	     \end{align}
	where we have introduced a would-be Higgs field $\phi$, a Dirac fermion $\psi$, which will later be chosen as a proxy for the electron, and the $U(1)$ gauge field $A_\mu$, which corresponds to the photon, with its usual field-strength tensor $F_{\mu\nu}=\partial_\mu A_\nu -\partial_\nu A_\mu$. 
	Note that the Dirac fermion is charged under $U(1)$, and it is minimally coupled to the photon field via the gauge covariant derivative
	\begin{equation}\label{eq:Dpsi}
	    D_\mu\psi=\partial_\mu\psi +i q A_\mu\psi,
	\end{equation}
	where $q$ is the electromagnetic coupling. 
	
	Before analysing the interactions induced by extending the gravitational sector beyond the usual Einstein--Hilbert action, we first need to insert all the minimal gravitational couplings that have so far been ignored by working in Minkowski spacetime. This means that, for every pair of contracted Lorentz indices, we must include a factor of the metric $g^{\mu\nu}$. Additionally, for every $\gamma$ matrix appearing in the Dirac Lagrangian, we must include a vierbein $e^\mu_a$, which satisfies
    $\eta^{a b}e^\mu_a e^\nu_b=g^{\mu\nu}$, where $\eta^{ab}$ is the flat spacetime metric. (We remind the reader that the flat-space indices of the vierbein are raised and lowered with the flat-space metric.) The latter is necessary since the algebra of the $\gamma$ matrices is defined with respect to the Minkowski metric, i.e., $\{\gamma^{a},\gamma^{
    b}\}=2\eta^{ab}$; the vierbeins relate the curved and flat, tangent spaces. By this means, we obtain the minimally coupled action
	\begin{align}\label{eq_QEDHiggs no CovDev}
	     S_{\rm m}[g_{\mu \nu}]=&\int \D^4{x} \sqrt{-g} \left[-\frac{1}{4}g^{\alpha\mu}g^{\beta\nu}F_{\alpha\beta}F_{\mu\nu}+\frac{1}{2}g^{\mu\nu}\partial_\mu \phi \partial_\nu \phi\right. \nonumber\\ 
	     &+i\bar{\psi}e^\mu_a\gamma^{a}\nabla_\mu\psi -y\bar{\psi}\phi\psi +\frac{1}{2}\mu^2 \phi^2 -\frac{\lambda}{4!}\phi^4 -\frac{3\mu^4}{2\lambda}\bigg],
	\end{align}
	where we have also included a factor of $\sqrt{-g}$ in the spacetime volume element. Herein, the Minkowski gauge covariant derivative has been promoted to the general covariant derivative. {We note that the Dirac-conjugate spinor is defined with respect to the flat-spacetime gamma matrix $\gamma_0$, i.e., $\bar{\psi}\equiv \psi^{\dag}\gamma_0$ (see, e.g., Ref.~\cite{Prinz:2020nru}).}

	For scalar fields, the gravitational covariant derivative just trivially reduces to a partial derivative, such that $\nabla_\mu\phi\to\partial_\mu\phi$. However, when acting on a \edit{covector} $Y_\rho$, the covariant derivative takes the form
	\begin{equation}
	    \nabla_\mu Y_\nu=\partial_\mu Y_\nu \edit{-} \Gamma^\rho_{\mu\nu} Y_\rho,
	\end{equation}
	where $\Gamma_{\mu\nu}^\rho=\frac{1}{2}g^{\rho\lambda}(\partial_{\mu}g_{\lambda\nu}+\partial_{\nu}g_{\mu\lambda}-\partial_{\lambda}g_{\mu\nu})$ are the usual Christoffel symbols. This definition for the covariant derivative is chosen such that $\nabla_\rho g_{\mu\nu}=0$, but it can take many other forms. For instance, we will later define and work with a different choice that will be more convenient for the specific case of Brans--Dicke theories~\cite{Copeland:2021qby}. However, \edit{as expected}, it does not matter which definition one uses in this action, given that the following property will always hold in QED:
	\begin{equation}
	    F_{\mu\nu}=\nabla_\mu A_\nu-\nabla_\nu A_\mu=\partial_\mu A_\nu-\partial_\nu A_\mu,
	\end{equation}
	since the curvature-dependent terms are symmetric under the permutation of $\mu$ and $\nu$. Finally, the covariant derivative acting on a fermion field, \edit{with its} dependence on the gauge field from QED, is given by
	\begin{equation}
	    \nabla_\mu\psi=\partial_\mu\psi +iq A_\mu\psi-\frac{i}{2}\Omega_\mu\psi,
	\end{equation}
	where 
	\begin{equation}
	    \Omega_\mu=(\Gamma_{ab})_\mu S^{ab}
	\end{equation}
	is the spin connection. The latter is defined by
	\begin{equation}
	    (\Gamma_{ab})_\mu=\edit{-}e_{a\alpha}e_b^\beta\Gamma_{\mu\beta}^\alpha+e_{a\alpha}\partial_\mu e_b^\alpha, \qquad \text{where} \qquad S^{ab}=\frac{i}{4}[\gamma^a,\gamma^b].
	\end{equation}
	With these minimal couplings now included, the action takes the form\footnote{\edit{Faddeev--Popov ghosts have been omitted, since they decouple in Abelian gauge theories in Minkowski spacetime. Note, however, that the ghost fields do not decouple from gravity, even in the Abelian case~\cite{Prinz:2020nru}. A full treatment of these terms in the context of non-Abelian gauge symmetries is provided in~\ref{appendix_ghosts}, wherein we describe the necessary inclusion of minimal gravitational couplings in the ghost Lagrangian.}}
	\begin{align}\label{ActionJordanFull}
	     S_{\rm m}[g_{\mu\nu}]=\int \D^4{x} \sqrt{-g} &\left[-\frac{1}{4}g^{\alpha\mu}g^{\beta\nu}F_{\alpha\beta}F_{\mu\nu}+\frac{1}{2}g^{\mu\nu}\partial_\mu \phi \partial_\nu \phi \right. \nonumber \\ 
	     &+i\bar{\psi}e^\mu_a\gamma^{a}\partial_\mu\psi +\frac{1}{2}\bar{\psi}e^\mu_a\gamma^{a}\Omega_\mu\psi-q\bar{\psi}e^\mu_a\gamma^{a}A_\mu\psi \nonumber \\
	     &-y\bar{\psi}\phi\psi +\frac{1}{2}\mu^2 \phi^2 -\frac{\lambda}{4!}\phi^4 -\frac{3\mu^4}{2\lambda}\bigg].
	\end{align}
	We can now proceed to append the gravitational sector.
	
	The minimal choice for the gravitational sector is the Einstein--Hilbert action, giving the full action
 \begin{equation}
		S=\int \D^4{x} \sqrt{-g} \left[-\frac{\Mpl^2}{2}R\right]+  S_{\rm m}[ g_{\mu \nu}],
		\label{ActionEinsteinHibert}
	\end{equation}
 where $R$ is the Ricci scalar, and $\Mpl$ is the Planck mass, which determines the strength of the gravitational force. We can, however, also consider extended gravitational sectors, and one of the simplest examples is the Brans--Dicke scalar-tensor theory~\cite{Brans:1961sx}, in which a dynamical scalar field replaces the Planck mass. Such theories are described by an action with the following generic form:
	\begin{equation}
		S=\int \D^4{x} \sqrt{-g} \left[-\frac{F(X)}{2}R + \frac{1}{2}Z(X)\partial_\mu X \partial^\mu X -U(X)\right]+  S_{\rm m}[ g_{\mu \nu}].
		\label{ActionGenericJordan}
	\end{equation}
	Herein, $X$ is a real scalar field, subject to the self-interaction potential $U(X)$ and coupled non-minimally to the Ricci scalar $R$ through the function $F(X)$. From a phenomenological perspective, there are tight constraints on the late-time evolution of Newton's gravitational ``constant'', e.g., from observations of the Moon's orbit~\cite{Muller:2007zzb}. We must therefore choose the functions $F(X)$, $Z(X)$ and $U(X)$, such that $\left<F(X)\right>=\Mpl^2$ is approximately constant, e.g., by $X$ obtaining an approximately constant vacuum expectation value (vev).
	Notice that the field $X$ is not or, at least, does not appear to be canonically normalized, by virtue of the function $Z(X)$ included in its kinetic term. In fact, additional contributions to the kinetic energy of the field $X$ arise through the coupling to the scalar curvature. Moreover, while the matter sector does not contain any direct couplings to the field $X$, these couplings may be hidden in the mixing between the tensor and scalar degrees of freedom of the extended gravitational sector. The interactions between the field $X$ and the would-be Standard Model fields become manifest once we have dealt with these mixings, and doing so is the main purpose of the package \code{FeynMG}.

For the Brans--Dicke example above, there are two ways that we can proceed, as will be described in the next subsections:
\begin{enumerate}
    \item We can make a Weyl rescaling of the metric to remove the non-minimal gravitational coupling of the field $X$ to the Ricci scalar, taking us to the so-called \textit{Einstein frame}.
    
    \item We can continue in the \textit{Jordan frame} (where the curvature couplings are manifest), by analysing how the metric degrees of freedom mediate interactions between the field $X$ and our would-be Standard Model fields.
    \end{enumerate}
    Before describing these two cases, however, it is important to note that the presence of additional non-minimal gravitational couplings, e.g., $R_{\mu\nu}\nabla^{\mu}\nabla^{\nu}X$ (as arises in the Horndeski class of scalar-tensor theories, where $R_{\mu\nu}$ is the Ricci tensor), the Weyl rescaling of the metric (or more generally a disformal transformation~\cite{Bekenstein:1992pj,Zumalacarregui:2013pma,Teixeira:2019hil} of the metric) may not be able to remove all non-minimal couplings simultaneously. In these cases, we may not be able to transform into an Einstein frame and will have little choice but to continue working with non-minimal interactions with gravity.


\subsection{Weyl transforming into the Einstein frame}
\label{Section_Calculation}

Our aim is to isolate the new interactions between the matter fields that arise because of the modifications to the gravitational sector. The most common way of doing this is to transform to the Einstein frame. This amounts to a redefinition of the curvature-dependent objects (called a Weyl \edit{or conformal} transformation) such that the resulting gravitational action does not present any non-minimal couplings.

For the Lagrangian defined in Eq.~\eqref{ActionJordanFull}, this transformation will take the following form
\begin{subequations}\label{eq-metric-transf}
	\begin{align}
	        g_{\mu\nu}\to &\frac{\tMpl^2}{F(X)}\tilde{g}_{\mu\nu}, &
	        g^{\mu\nu}\to &\frac{F(X)}{\tMpl^2}\tilde{g}^{\mu\nu},\\
	        e_{a}^{\mu}\to &\frac{\tMpl}{\sqrt{F(X)}}\tilde{e}_{a}^{\mu}, &
	        e^{a}_{\mu}\to &\frac{\sqrt{F(X)}}{\tMpl}\tilde{e}^{a}_{\mu},
	\end{align}
	\end{subequations}
	where $\tilde{g}_{\mu\nu}$, $\tilde{e}_a^\mu$ and $\tMpl$ are the metric, vierbein and Planck mass in the Einstein frame, respectively. \edit{These transformations are only well-defined for $F(X)>0$, which is satisfied as long as the effective $\Mpl^2$ is positive in the Jordan frame.} To get through the algebra, the following transformations will be useful:
	\begin{subequations}\label{Ricci_transformations}
	\begin{align}
	        \sqrt{-g}\frac{F(X)}{2}R\to & \sqrt{-\tilde{g}}\left(\frac{\tMpl^2}{2}\tilde{R} -\frac{3\tMpl^2F'(X)^2}{4F(X)^2}\tilde{g}^{\mu\nu}\partial_\mu X\partial_\nu X\right),\\
	       e_a^\mu\gamma^a \Omega_\mu\to& \frac{\sqrt{X}}{\tMpl}\tilde{e}_a^\mu\gamma^a\left(\tilde{\Omega}_\mu - \frac{3i}{2}\frac{F'(X)}{F(X)}\partial_\mu X\right),
	\end{align}
	\end{subequations}
	where $F'(X)={\partial}F(X)/{\partial}X$ and all the curvature-dependent quantities with a tilde are built with the Einstein-frame metric $\tilde{g}_{\mu\nu}$ or vierbein $\tilde{e}_a^\mu$.
	
	Applying the transformations in Eq.~\eqref{eq-metric-transf} to the Jordan-frame action, we obtain
	\begin{align}\label{ActionEinsteinFull}
	     S=&\int \D^4{x} \sqrt{-\tilde{g}} \left[-\frac{\tMpl^2}{2}R + \frac{\tMpl^2}{2}\left[\frac{Z(X)}{F(X)}+\frac{3F'(X)^2}{2F(X)^2}\right]\tilde{g}^{\mu\nu}\partial_\mu X \partial_\nu X \right.\nonumber\\ &-\frac{1}{4}\tilde{g}^{\alpha\mu}\tilde{g}^{\beta\nu}F_{\alpha\beta}F_{\mu\nu}+\frac{\tMpl^2}{2F(X)}\tilde{g}^{\mu\nu}\partial_\mu \phi \partial_\nu \phi-q\frac{\tMpl^3}{F(X)^{3/2}}\bar{\psi}\tilde{e}^\mu_a\gamma^{a}A_\mu\psi\nonumber \\ 
	     &+i\frac{\tMpl^3}{F(X)^{3/2}}\bar{\psi}\tilde{e}^\mu_a\gamma^{a}\partial_\mu\psi +\frac{1}{2}\frac{\tMpl^3}{F(X)^{3/2}}\bar{\psi}\tilde{e}^\mu_a\gamma^{a}\psi\left(\tilde{\Omega}_\mu-\frac{3i}{2}\frac{F'(X)}{F(X)}\partial_\mu X\right) \nonumber\\
	     &-\frac{\tMpl^4}{F(X)^2}\left(y\bar{\psi}\phi\psi -\frac{1}{2}\mu^2 \phi^2 +\frac{\lambda}{4!}\phi^4 +\frac{3\mu^4}{2\lambda}\right)-\frac{\tMpl^4}{F(X)^2}U(X)\bigg],
	\end{align}
	wherein we have recovered a canonical Einstein--Hilbert term for the gravitational action. However, all the couplings of the Brans--Dicke scalar arising from the modification of gravity now appear explicitly in the matter Lagrangian. Notice, in particular, that most of the kinetic energies of the fields are not canonically normalised due to these new couplings.
	
	To canonically normalise the field $X$, we must solve the integral
	\begin{equation}
	    \tilde{X}(X)\equiv\tMpl\int^X_{X_0}\D\hat{X}\sqrt{\frac{Z(\hat{X})}{F(\hat{X})}+\frac{3F'(\hat{X})^2}{2F(\hat{X})^2}},
	\end{equation}
	where $X_0$ is taken to be zero for simplicity. For the rest of the fields, we rescale them according to their classical scaling dimension, i.e.,
	\begin{align}
	    \psi\to&\sqrt{\frac{\tilde{F}(\tilde{X})^{3/2}}{\tMpl^3}}\tilde{\psi}, & 	     \phi\to&\frac{\sqrt{\tilde{F}(\tilde{X})}}{\tMpl}\tilde{\phi},
	\end{align}
	where $\tilde{F}(\tilde{X})\equiv F(X)$. With this, the Lagrangian takes the following form:
	\begin{align}\label{ActionEinsteinFullRescaled}
	     \lgr=&-\frac{\tMpl^2}{2}R + \frac{1}{2}\tilde{g}^{\mu\nu}\partial_\mu \tilde{X} \partial_\nu \tilde{X}-\frac{1}{4}\tilde{g}^{\alpha\mu}\tilde{g}^{\beta\nu}F_{\alpha\beta}F_{\mu\nu}\nonumber \\ 
	     &+i\bar{\tilde{\psi}}\tilde{e}^\mu_a\gamma^{a}\partial_\mu\tilde{\psi} +\frac{1}{2}\bar{\tilde{\psi}}\tilde{e}^\mu_a\gamma^{a}\tilde{\Omega}_\mu\tilde{\psi}-q\bar{\tilde{\psi}}\tilde{e}^\mu_a\gamma^{a}A_\mu\tilde{\psi}\nonumber\\
	     &+\frac{1}{2}\tilde{g}^{\mu\nu}\partial_\mu\tilde\phi \partial_\nu \tilde\phi+\frac{1}{2}\frac{\tilde{F}'(\tilde{X})}{\tilde{F}(\tilde{X})}\tilde{\phi}\tilde{g}^{\mu\nu}\partial_\mu\tilde\phi \partial_\nu \tilde{X}\nonumber\\
	     &+\frac{1}{8}\left(\frac{\tilde{F}'(\tilde{X})}{\tilde{F}(\tilde{X})}\right)^2\tilde{\phi}^2\tilde{g}^{\mu\nu}\partial_\mu\tilde{X} \partial_\nu \tilde{X}-y\bar{\tilde\psi}\tilde\phi\tilde\psi\nonumber\\
	     &+\frac{\tMpl^2}{\tilde{F}(\tilde{X})}\frac{1}{2}\mu^2\tilde{\phi}^2 -\frac{\lambda}{4!}\tilde{\phi}^4 -\frac{3\mu^4}{2\lambda}\frac{\tMpl^4}{\tilde{F}(\tilde{X})^2}-\frac{\tMpl^4}{\tilde{F}(\tilde{X})^2}\tilde{U}(\tilde{X}),
	\end{align}
	where $\tilde{U}(\tilde{X})\equiv U(X)$ and $\tilde{F}'(\tilde{X})=\partial \tilde{F}(\tilde{X})/\partial\tilde{X}$. Thus, one of the main inconveniences of working in the Einstein frame is that it loses the simplicity of the Lagrangian defined in the Jordan frame. This is because the Weyl transformation and the redefinition of the fields introduces factors of $\tilde{F}(\tilde{X})$ throughout the Lagrangian, which, on making a series expansion of $\tilde{F}(\tilde{X})$, will introduce infinite towers of operators that involve the SM fields and increasing powers of the scalar field $\tilde{X}$.
	
	At this point, we can already make an important observation:~The couplings between the SM fields and the scalar field $\tilde{X}$ arise only through the scalar kinetic terms and terms with dimensionful parameters, i.e., those terms that are not invariant under Weyl transformations. Thus, for the Standard Model (illustrated already by the toy model described here), the modifications to the dynamics from the new scalar field $\tilde{X}$ are, in the Einstein frame, communicated by the Higgs sector, with the squared mass parameter $\mu^2$ of the tree-level Higgs potential playing the dominant role at low momentum exchange. In this way, there are strong parallels between the Brans--Dicke-type scalar-tensor theories and Higgs-portal theories (see Ref.~\cite{Burrage:2018dvt}).
	
	Expanding the fields around their vacuum expectation values will give rise to kinetic and mass mixings between $\tilde{\phi}$ and $\tilde{X}$. Thus, when two fermions interact via their Yukawa coupling and exchange a would-be Higgs boson ($\tilde{\phi}$) in the $t$ channel, there are two contributions to the central potential:~a short-range interaction due to the heavy mode (the Higgs boson) and a long-range interaction due to the light mode (the light, additional scalar boson), see Ref.~\cite{Burrage:2018dvt}. Such long-range forces arising from the additional scalar fields of extended gravity sectors are often referred to as ``fifth forces''. In this way, even if the original matter Lagrangian is only minimally coupled to gravity in the Jordan frame, there can be experimentally testable modifications to the force laws that depend on the dynamics of the new scalar field.
	
	Given how these new interactions manifest in the Einstein frame, it is instructive to consider how the same modifications to the dynamics manifest in the Jordan frame, without making the Weyl transformation (at least at first). This is the focus of the next subsection.
	

\subsection{Staying in the Jordan frame}
\label{Subsection Calculation Jordan}

    We can determine the modifications to the dynamics without performing a Weyl transformation to the Einstein frame and work directly in the Jordan frame. In this frame, new interactions between the fields of the matter sector arise through the gravity sector itself, and we proceed by perturbing the metric around a flat spacetime~\cite{Fierz:1939ix,Donoghue:1995cz,Donoghue:2017pgk} in the gravitational weak-field limit.
    
    Expanding the metric up to leading order in perturbations corresponds to
    \begin{subequations}
        \label{eq:expansion}
	\begin{align}
	    g_{\mu\nu}&=\eta_{\mu\nu} + h_{\mu\nu},\\
	    g^{\mu\nu}&=\eta^{\mu\nu} - h^{\mu\nu}\edit{+\cdots},	        
	\end{align}
	\end{subequations}
	where $\eta_{\mu\nu}$ is the usual flat spacetime metric and $h_{\mu\nu}$ is the perturbation in the metric, which, once quantized, corresponds to the graviton. The higher-order terms in the expansion of $g_{\mu\nu}$ are necessary to satisfy $g_{\mu\nu}g^{\nu\rho}=\delta^{\rho}_{\mu}$ to all orders.
	
	For the gravitational sector of the Brans--Dicke-like theory [Eq.~\eqref{ActionGenericJordan}], with action
	\begin{equation}
	    S_{\rm G}=\int \D^4{x} \sqrt{-g} \left[-\frac{F(X)}{2}R \right],
	\end{equation}
	we obtain the following expansion up to second order in the fields:
	\begin{align}\label{Eq_Jordan_nogauge}
	    \lgr_{\rm{G}}=&\frac{F(X)}{2}\left(\frac{1}{4}\partial_\rho h_{\mu\nu}\partial^\rho h^{\mu\nu} -\frac{1}{4} \partial_\mu h\partial^\mu h +\frac{1}{2}\partial^\mu h_{\mu\nu}\partial^\nu h -\frac{1}{2}\partial^\mu h_{\mu\rho}\partial_\nu h^{\nu\rho}\right)\nonumber\\
	    &-\frac{F'(X)}{4} \partial_\mu X \partial^\mu h+\frac{F'(X)}{2}\partial_\mu X \partial_\nu h^{\mu\nu}.
	\end{align}
	
	It still remains to fix a gauge, and one choice is the harmonic gauge, which satisfies the following condition:
	\begin{equation}
	    \nabla_\mu\nabla^\mu=\partial_\mu\partial^\mu\to g^{\mu\nu}\Gamma^\rho_{\mu\nu}=0.
	\end{equation}
	This can be introduced into the Lagrangian through the term
	\begin{equation}
	    \lgr_{\rm GF}=\frac{\Mpl^2\xi}{4}g_{\mu\nu}\Gamma^\mu\Gamma^\nu,
	\end{equation}
	where $\Gamma^\mu=g^{\alpha\beta}\Gamma^\mu_{\alpha\beta}$ \edit{and $\xi$ is a Lagrange multiplier that we will set to $\xi=1$}. With this gauge choice, linearization of Einstein--Hilbert gravity leads to the familiar Fierz--Pauli Lagrangian~\cite{Fierz:1939ix}, given by 
	 \begin{equation}\label{Eq_FiewzP}
	    \lgr_{\rm{FP}}=\frac{\Mpl^2}{4}\left(\frac{1}{2}\partial_\rho h_{\mu\nu}\partial^\rho h^{\mu\nu}-\frac{1}{4} \partial_\mu h\partial^\mu h  \right).
	\end{equation}
	
	When working with Brans--Dicke theories in the Jordan frame, it is convenient to use a different gauge:\ one that maps to the harmonic gauge when performing the Weyl transformation to the Einstein frame.\footnote{This argument can also be applied when gauge fixing other fields. See \ref{appendix_A} for a demonstration of working with the U(1) gauge field.} This can be achieved by redefining the covariant derivative such that its action on a \edit{covector} $Y_\nu$ is as follows:
	\begin{equation}\label{eq_ModifiedCovDev}
	    \mathcal{D}_\mu Y_\nu= \partial_\mu Y_\nu \edit{-} \Gamma^\rho_{\mu\nu}Y_\rho \edit{-} C^\rho_{\mu\nu} Y_\rho,
	\end{equation}
	where
	\begin{equation}
	    C^\rho_{\mu\nu}=\frac{F'(X)}{2F(X)}(\delta^\rho_\mu\partial_\nu X+\delta^\rho_\nu\partial_\mu X - g_{\mu\nu}\partial^{\rho} X).
	\end{equation}
	This modified covariant derivative will map to $\nabla_\mu$ when going to the Einstein frame and satisfies the identity $\mathcal{D}_\rho(F(X)g_{\mu\nu})=0$ while preserving diffeomorphism invariance in the action, as shown in Ref.~\cite{Copeland:2021qby, DeWitt:1964mxt, Barvinsky:1985an, Finn:2022rlo}. We can then define a {\it scalar-harmonic gauge} condition in terms of the new covariant derivative, namely
	\begin{equation}
	    \mathcal{D}^\mu \mathcal{D}_\mu=\partial^\mu \partial_\mu \to g^{\mu\nu} \Gamma^\rho_{\mu\nu} -\frac{F'(X)}{F(X)}\partial^\rho X=0. 
	\end{equation}
	This can be introduced into the Lagrangian as
	\begin{equation}\label{GFterm}
	\lgr_{\rm{GF}}=\frac{F(X)\xi}{4}g_{\alpha\beta}\left[g^{\mu\nu}\Gamma^\alpha_{\mu\nu} -\frac{F'(X)}{F(X)}\partial^\alpha X\right]\left[g^{\sigma\rho} \Gamma^\beta_{\sigma\rho} -\frac{F'(X)}{F(X)}\partial^\beta X\right],
	\end{equation}
	\edit{where we have introduced the explicit coupling to the Lagrange multiplier $\xi$, which will again be set to $\xi=1$. }Expanding this gauge fixing term around a Minkowski background and adding it to the linearized gravitational sector from Eq.~\eqref{Eq_Jordan_nogauge}, we obtain
	\begin{align}\label{LagGauge}
	    \lgr=&\frac{F(X)}{4}\left(\frac{1}{2}\partial_\rho h_{\mu\nu}\partial^\rho h^{\mu\nu}-\frac{1}{4} \partial_\mu h\partial^\mu h\right)\nonumber\\
	    +&\frac{F'(X)^2}{4F(X)}\partial_\mu X \partial^\mu X -\frac{F'(X)}{4} \partial_\mu X \partial^\mu h.
	\end{align}
	Herein, we have recovered the usual kinetic energy terms of the graviton, as appear in the Fierz--Pauli Lagrangian~\eqref{Eq_FiewzP}, with the exception that non-minimal couplings to the field $X$ appear through the overall factor of $F(X)$. Notice that the Lagrangian~\eqref{LagGauge} contains two additional terms relative to the Fierz--Pauli Lagrangian~\eqref{Eq_FiewzP}.  The first contributes to the kinetic energy of the field $X$, which will have to be canonically normalized, and the second is a kinetic interaction between $X$ and the trace of the graviton $h$. As we will show later, it \edit{is} this kinetic mixing that leads to additional interactions between the matter fields.
	
    On including the matter sector from the original action in Eq.~\eqref{ActionJordanFull}, we obtain the following Lagrangian after the linearization up to first order in $1/\sqrt{F(X)}$ (noting that $\Mpl^2=F(v_X)$): 
    \begin{align}\label{eq_Jrd_notcanonX}
	    \lgr=&\frac{F(X)}{4}\left(\frac{1}{2}\partial_\rho h_{\mu\nu}\partial^\rho h^{\mu\nu}-\frac{1}{4} \partial_\mu h\partial^\mu h\right)-\frac{F'(X)}{4} \partial_\mu X \partial^\mu h \nonumber\\
	    -&\frac{1}{4}F_{\mu\nu}F^{\mu\nu}+\left[\frac{Z(X)}{2}+\frac{F'(X)^2}{4F(X)}\right]\partial_\mu X \partial^\mu X +\frac{1}{2}\partial_\mu \phi \partial^\mu \phi\nonumber\\+&i\bar{\psi}\gamma^{\mu}\partial_\mu\psi -q\bar{\psi}\gamma^{\mu}A_\mu\psi -y\bar{\psi}\phi\psi \nonumber\\
	    +&\frac{1}{2}\mu^2 \phi^2 -\frac{\lambda}{4!}\phi^4 -\frac{3\mu^4}{2\lambda}-U(X)+\frac{1}{2}h_{\mu\nu}T^{\mu\nu}\bigg]+\cdots,
	\end{align}
	where graviton self-interactions have been ignored and $T_{\mu\nu}$ is the energy-momentum tensor of the matter sector.
	
	The kinetic energy of the $X$ field can be canonically normalized by defining
	\begin{equation}\label{eq_normX}
	    \chi(X)=\int_{X_0}^{X}\D\hat{X}\sqrt{Z(\hat{X})+\frac{F'(\hat{X})^2}{2F(\hat{X})}},
	\end{equation}
	where $X_0$ is again taken to be zero for simplicity. Doing so leads to the Lagrangian 
	\begin{align}
	    \lgr_{\rm G}=&\frac{\hat{F}(\chi)}{4}\left(\frac{1}{2}\partial_\rho h_{\mu\nu}\partial^\rho h^{\mu\nu}-\frac{1}{4} \partial_\mu h\partial^\mu h\right)-\frac{\hat{F}'(\chi)}{4} \partial_\mu \chi \partial^\mu h \nonumber\\
	    -&\frac{1}{4}F_{\mu\nu}F^{\mu\nu}+\frac{1}{2}\partial_\mu \chi \partial^\mu \chi +\frac{1}{2}\partial_\mu \phi \partial^\mu \phi\nonumber\\
	    +&i\bar{\psi}\gamma^{\mu}\partial_\mu\psi-q\bar{\psi}\gamma^{\mu}A_\mu\psi -y\bar{\psi}\phi\psi\nonumber \\
	    +&\frac{1}{2}\mu^2 \phi^2 -\frac{\lambda}{4!}\phi^4 -\frac{3\mu^4}{2\lambda}-\hat{U}(\chi)+\frac{1}{2}h_{\mu\nu}T^{\mu\nu}\bigg]+\cdots,
	\end{align}
	where $\hat{F}(\chi)\equiv F(X)$, $\hat{F}'(\chi)= \partial \hat{F}(\chi)/\partial \chi$ and $\hat{U}(\chi)\equiv U(X)$. Now, we have only the graviton left to canonically normalise, since it is still non-minimally coupled to the function $\hat{F}(\chi)$. However, as noted previously, the potential
	$\hat{U}(\chi)$ must lead to a non-vanishing vacuum expectation value for $\chi$ at late times so that the theory mimics Einstein gravity.\footnote{We might expect $v_{\chi}$ to be evolving on cosmological timescales, but these timescales are long compared to the those relevant for elementary particle interactions.} With this in mind, we shift $\chi\to\chi+v_\chi$ to obtain
    \begin{align}\label{eq_Jordan_mixing}
	    \lgr=& \frac{\hat{F}(v_\chi)}{4}\left(\frac{1}{2}\partial_\rho h_{\mu\nu}\partial^\rho h^{\mu\nu}-\frac{1}{4} \partial_\mu h\partial^\mu h\right)-\frac{\hat{F}'(v_\chi)}{4} \partial_\mu \chi \partial^\mu h \nonumber\\
	    -&\frac{1}{4}F_{\mu\nu}F^{\mu\nu}+\frac{1}{2}\partial_\mu \chi \partial^\mu \chi +\frac{1}{2}\partial_\mu \phi \partial^\mu \phi\nonumber\\+&i\bar{\psi}\gamma^{\mu}\partial_\mu\psi -q\bar{\psi}\gamma^{\mu}A_\mu\psi -y\bar{\psi}\phi\psi\nonumber\\
	    +&\frac{1}{2}\mu^2 \phi^2 -\frac{\lambda}{4!}\phi^4 -\frac{3\mu^4}{2\lambda} -\hat{U}(\chi+v_\chi) +\frac{1}{2}h_{\mu\nu}T^{\mu\nu}+\cdots,
	\end{align}
	where higher-order terms in the interactions between $\chi$ and $h_{\mu\nu}$ have been omitted in the ellipsis.
	
\begin{figure}
	    \centering
	    \includegraphics[scale=0.4]{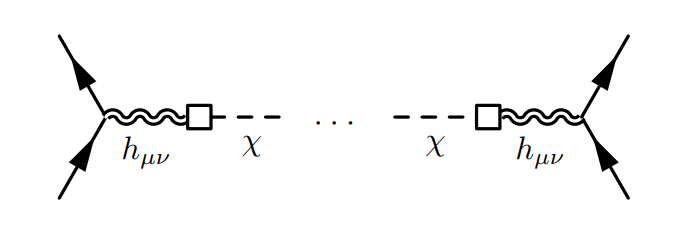}
	    \caption{Series of diagrams contributing to the fifth force, and arising from the kinetic mixing between the graviton $h_{\mu\nu}$ and the scalar field $\chi$. The ellipsis represents the series summing over all insertions of the kinetic mixing.}
	    \label{fig:fifth}
	\end{figure}

	The modification of gravity leads to a kinetic mixing between the trace of the graviton $h$ and the $\chi$ field; the last term in the first line of Eq.~\eqref{eq_Jordan_mixing}. The example of the fifth-force exchange described in the previous section then manifests in the Jordan frame through this mixing, as shown in Figure \ref{fig:fifth} (see Ref.~\cite{Copeland:2021qby}).
	
	We can remove this mixing by the following transformation of the graviton and $\chi$ field:\footnote{This transformation can be calculated in multiple ways. In Ref.~\cite{Copeland:2021qby}, this was achieved by solving the system of equations that left the action diagonalized. Alternatively, in \ref{appendix_C}, we show how to achieve this directly by transforming the kinetic matrix.}
	\begin{subequations}\label{eq_diagonalization_h_chi}
	\begin{align}
	h_{\mu\nu} \to & \frac{2}{\Mpl} h_{\mu\nu} +\frac{1}{\Mpl} \frac{\hat{F}'(v_\chi)}{\sqrt{\Mpl^2+\hat{F}'(v_\chi)^2}}\sigma \eta_{\mu\nu}, \\
	\chi \to & -\frac{1}{\sqrt{1+\left(\frac{\hat{F}'(v_\chi)}{\Mpl}\right)^2}}\sigma,
	\end{align}
	\end{subequations}
	where $\hat{F}(v_\chi)= \Mpl^2$ has been substituted and $\sigma$ corresponds to the canonically normalized scalar field. This amounts to a perturbative implementation of the Weyl transformation, as is clear when one considers the resulting Lagrangian
	\begin{align}\label{eq_Jordan_fin}
	    \lgr=&\frac{1}{4} \partial_\mu h\partial^\mu h-\frac{1}{2}\partial_\rho h_{\mu\nu}\partial^\rho h^{\mu\nu}-\frac{1}{4}F_{\mu\nu}F^{\mu\nu}\nonumber\\
	    +&i\bar{\psi}\gamma^{\mu}\partial_\mu\psi -q\bar{\psi}\gamma^{\mu}A_\mu\psi -y\bar{\psi}\phi\psi-\hat{U}(\chi(\sigma)+v_\chi)\nonumber\\
	    +&\frac{1}{2}\partial_\mu \sigma \partial^\mu \sigma +\frac{1}{2}\partial_\mu \phi \partial^\mu \phi+\frac{1}{2}\mu^2 \phi^2 -\frac{\lambda}{4!}\phi^4 -\frac{3\mu^4}{2\lambda}\nonumber\\
	    +&\frac{1}{\Mpl}h_{\mu\nu}T^{\mu\nu}+\frac{1}{2\Mpl}\frac{\hat{F}'(v_\chi)}{\sqrt{\Mpl^2+\hat{F}'(v_\chi)^2}}\sigma T_\mu^\mu+\cdots,
	\end{align}
	where $T_\mu^\mu$ is the trace of the matter energy-momentum tensor. The fifth force arising from the final term in Eq.~\eqref{eq_Jordan_fin} will depend on the trace of the energy-momentum tensor of the interacting particles, leading to at most derivative interactions with $\sigma$ for scale-invariant sectors~\cite{Ferreira:2016kxi} \edit{(cf. Ref. \cite{Freese}, wherein the energy-momentum tensors for the different sectors of the SM are given, including ghost fields)}.

	 We have seen that working in the Jordan frame requires us to linearize the gravitational sector and to diagonalize the fields, while, in the Einstein frame, we had to perform the Weyl transformation and various rescalings of the matter fields, losing the simplicity of the potentials in the process. Whichever approach we take, the overall message of this section is not a discussion on which frame is best for calculations, as it is a matter of preference, but the fact that deriving Feynman rules for scalar-tensor theories is a tedious and time-consuming task, even for the simplest models. 
	
	This begs for a tool that helps us automate this process. In the rest of this paper, we will introduce the \code{Mathematica} package {\code{FeynMG}} within the \code{FeynRules} environment, which can efficiently perform manipulations on scalar-tensor theories of the types described in this section.
	

\section{Implementation} 
\label{sec:implementation}

	\code{FeynMG} implements calculations of the type described in Section \ref{sec:method}. The only necessary input is a model file compatible with \code{FeynRules} containing the matter Lagrangian and the description of all the existing fields and parameters. The user can then supplement this Lagrangian with their chosen scalar-tensor theory.
	
	Scalar-tensor theories will generally give rise to both mass and kinetic mixings between fields. While FeynRules can deal with mass mixing if pre-defined in the model file, it cannot deal with kinetic mixing or cases where the form of the mass mixing is not known a priori.  This is because \code{FeynRules} will ignore terms of quadratic order and will assume that all fields are canonically normalized. The scope of \code{FeynMG} is to linearize gravity and perform the necessary redefinitions to the fields such that it can be consistently used by \code{FeynRules} and all compatible packages.
	
	We aim to make the code as easy to use as possible without losing the generality in the model files and desired gravitational actions. For example, for the input Lagrangian, it is possible to use an action defined in flat spacetime (i.e., reuse a \code{FeynRules} model file without modifying it). This is possible thanks to the function \mmaInlineCell{Input}{\mmaDef{InsertCurv}}, which for every pair of contracted indices will insert a metric $g^{\mu\nu}$ or vierbein $e^{\mu a}$, as appropriate, and promote partial derivatives to covariant derivatives.
	
	Once all the minimal curvature dependencies are inserted into the Lagrangian, we need to append a gravitational action, wherein, e.g., the Ricci scalar can be specified using \mmaInlineCell{Input}{\mmaDef{RScalar}} (see \ref{appendix:functions:curvature} for the list of defined curvature objects). As is the case for \code{FeynRules}, it is necessary to identify fields and parameters. These attributes can be assigned to variables by using the functions \mmaInlineCell{Input}{\mmaDef{AddScalar[]}} and \mmaInlineCell{Input}{\mmaDef{AddParameter[]}}, respectively, allowing complete freedom when creating the gravitational sector, and any number of new scalar degrees of freedom and parameters to be defined. In principle, the package should be able to deal with any gravitational sector, but it becomes more complicated the further we go beyond Brans--Dicke theories. The effective Planck mass can be extracted at any point in the calculation by using the function \mmaInlineCell{Input}{\mmaDef{GiveMpl}}. Moreover, using \mmaInlineCell{Input}{\mmaDef{InsertMpl}} will calculate the effective $\Mpl$ from the action and substitute it into the expression.
	
	As shown previously, in the particular case of Brans--Dicke gravity, we can perform a Weyl transformation such that the gravitational sector is of Einstein--Hilbert form and the matter action is instead dressed with additional scalar interactions. This is implemented in \code{FeynMG} by the function \mmaInlineCell{Input}{\mmaDef{ToEinsteinFrame}}. However, more general scalar-tensor theories may not have an Einstein frame, forcing us to stay in the Jordan frame and proceed by linearizing gravity. The latter is implemented by the function \mmaInlineCell{Input}{\mmaDef{LinearizeGravity}}, where the gravitational sector will be expanded up to second order, generating the kinetic energy for the graviton, and the matter sector will be expanded up to linear order in the interactions with the metric perturbation $h_{\mu\nu}$. Moreover, the Jacobian $\sqrt{-g}$ will be automatically inserted, unless the option \mmaInlineCell{Input}{\{\mmaDef{Jacobian->0ff}\}} is provided. 
	
	As described in the previous section, in the case of Brans--Dicke-like theories, it can be convenient to use the scalar-harmonic gauge from Eq.~\eqref{GFterm}. By specifying the option \mmaInlineCell{Input}{\{\mmaDef{SHGauge->0n}\}},  \mmaInlineCell{Input}{\mmaDef{LinearizeGravity}} will determine the scalar-harmonic gauge fixing term and append it to the Lagrangian, depending on the specific coupling function $F(X)$.\footnote{Similarly, it is possible to update the rest of the covariant derivatives in the Lagrangian to their modified forms by specifying the option \code{\{{UpdDevs->0n}\}}.}	This gauge choice will likely leave \mmaInlineCell{Input}{\mmaDef{CMod}} terms in the linearized Lagrangian, corresponding to the modification of the Christoffel symbols
	\begin{equation}\label{CmodImpl}
	    C^\rho_{\mu\nu}=\frac{F'(X)}{2F(X)}(\delta^\rho_\mu\partial_\nu X+\delta^\rho_\nu\partial_\mu X - g_{\mu\nu}\partial^{\rho} X).
	\end{equation}
	 Notice that the $\frac{F'(X)}{2F(X)}$ prefactor will have to be expanded in terms of $X$. Once this expansion is truncated at some order in $X$, we can no longer make a non-linear redefinition of the $X$ field (such as $X\to X^2$), since the ignored higher-order terms will give contributions at lower orders. To avoid this problem, \mmaInlineCell{Input}{\mmaDef{CMod}} won't be expanded until all the kinetic energies of the scalar fields have been canonicalized.
	 
	 When dealing with tensor algebra, we are used to working with Einstein's index notation, for which the following holds:~$A_\mu A^\mu=A_\rho A^\rho$. However, \code{Mathema-} \code{tica} will treat both terms $A_\mu A^\mu$ and $A_\rho A^\rho$ as distinct, since their indices are not represented by the same variable, leading to an overly complicated and long expression filled with repeated terms. The function \mmaInlineCell{Input}{\mmaDef{IndexSimplify}} deals with this problem by replacing indices term by term from a user-supplied set of indices, so that the expression can be simplified using \code{Mathematica}'s native functionality. 
	
	 From here, which frame we use is unimportant, since the package has all the tools to leave the Lagrangian in a form that is readable by \code{FeynRules}. If we stay in the Jordan frame (as may be necessary for theories that do not have an Einstein frame), one first needs to normalize the fields canonically. For scalar fields, the canonical normalization is implemented by the function \mmaInlineCell{Input}{\mmaDef{CanonScalar}}, which will find and normalize the lowest-order derivative term of every field. In the case where the lowest order is already very complicated, one can use the in-built \code{Mathematica} function \mmaInlineCell{Input}{\mmaDef{Series}} to perform a series expansion. 
	
	Similarly, we also need to normalize the graviton kinetic energy canonically. For that, depending upon the gravitational action, we might need to expand the fields around their vacuum expectation values, using \mmaInlineCell{Input}{\mmaDef{VevExpand}}, which first calculates all the possible values for the vevs, and then shifts all the fields around the user's chosen branch of solutions. Once the graviton kinetic energy has a constant prefactor, we can then use \mmaInlineCell{Input}{\mmaDef{CanonGravity}}, leaving all the fields canonically normalized with derivative interactions. As mentioned before, it will be at this point where all the \mmaInlineCell{Input}{\mmaDef{CMod}} terms arising from the modified covariant derivatives will be expanded to make manifest their dependence on the additional scalar degree of freedom arising from the extended gravitational sector. 
	
	The only thing left to do is to deal with any mass or kinetic mixings that have arisen between any of the metric and scalar degrees of freedom. As mentioned previously, \code{FeynRules} assumes that all fields are canonical and only works with terms higher than quadratic order, so any mixing terms in the Lagrangian would be ignored in the outcome. To deal with this, \mmaInlineCell{Input}{\mmaDef{MassDiagMG}} or \mmaInlineCell{Input}{\mmaDef{KineticDiagMG}} diagonalizes the scalar field masses or kinetic energies, respectively. When proceeding in the Jordan frame, as we saw in the last section, the dominant modifications to the dynamics arise through kinetic mixing between the additional scalar field and the trace of the graviton (cf., e.g., Figure \ref{fig:fifth}). The function \mmaInlineCell{Input}{\mmaDef{GravKinMixing}} will calculate and substitute into the Lagrangian the field redefinitions that diagonalizes this kinetic mixing, i.e., the equivalent of Eq.~\eqref{eq_diagonalization_h_chi}. With this, the Lagrangian \edit{is} in a form ready to be used by \code{FeynRules}. 

    Linearizing gravity and manipulating the Lagrangian into a form amenable to \code{FeynRules} can take significant computing time for extensive or complicated models. So that this process does not need to be repeated each time, the user can use the function \mmaInlineCell{Input}{\mmaDef{OutputModelMG}} to create a new model file from the final form of the Lagrangian produced by \code{FeynMG}, which includes all the information about the redefined fields, the parameters of the extended model and the effective Lagrangian itself. This model file can then be used directly in \code{FeynRules} without the need to rerun the manipulations implemented by \code{FeynMG}.
    
    To summarize, the package \code{FeynMG} provides a set of tools to help the user to upgrade the original \code{FeynRules} model file to one that includes the degrees of freedom of a canonical or extended gravitational sector.
    

\section{Usage}
\label{sec:usage}

	In this section, we provide the instructions for loading \code{FeynMG} and use it to perform the manipulations described in the preceding sections. We will work in the Jordan Frame, given that the same tools can be used for the Einstein frame. In \ref{appendix:functions}, we provide a summary of the tools provided by \code{FeynMG}. 


\subsection{Installation}

	\code{FeynMG} has dependencies on \code{FeynRules}, so both packages need to be loaded into \code{Mathematica} to make use of \code{FeynMG}. This can be done by running 
\begin{mmaCell}{Input}
  \mmaDef{<< {FeynRules}\` }
  \mmaDef{<< FeynMG\` }
\end{mmaCell}
within the appropriate working directory (set via \mmaInlineCell{Input}{\mmaDef{SetDirectory}[]}).

The next step is to load a model file that is compatible with \code{FeynRules} using the \code{FeynRules} function \mmaInlineCell{Input}{\mmaDef{LoadModel}[]} (for an extensive description of how to build these files, see Ref.~\cite{feynrules}). As mentioned previously, this model file does not need to include gravity in the defined fields or Lagrangians; these can be appended through \code{FeynMG}, as described earlier in Section \ref{sec:implementation}.


\subsection{Defining a gravitational action and transforming to the Einstein Frame}

Throughout this section, we will work with the same Lagrangian from Eq.~\eqref{ActionJordanFull}, whose matter sector is defined via
	\begin{align}
	     \code{LQED}=& -\frac{1}{4}F_{\mu\nu}F^{\mu\nu}+\frac{1}{2}\partial_\mu \phi \partial^\mu \phi -\alpha \partial^\mu A_\mu\partial^\nu A_\nu\nonumber\\ 
	     &+i\bar{\psi}\gamma^{\mu}\partial_\mu\psi -q\bar{\psi}\gamma^{\mu}A_\mu\psi -y\bar{\psi}\phi\psi\nonumber\\
	     & +\frac{1}{2}\mu^2 \phi^2 -\frac{\lambda}{4!}\phi^4 -\frac{3\mu^4}{2\lambda}.
	\end{align}
Note that the last term of the first line corresponds to the \edit{covariant} gauge-fixing term for the U(1) gauge field. 

The first thing to do is to introduce the minimal gravitational couplings of this matter Lagrangian. This amounts to inserting metrics or vierbeins, as appropriate, for each pair of contracted indies, and promoting all partial derivatives to covariant ones. To implement this in \code{FeynMG}, we run
\begin{mmaCell}[functionlocal=y]{Code}
  LCurv=\mmaDef{InsertCurv}[\mmaDef{LQED}]
\end{mmaCell}
{\footnotesize 
\begin{mmaCell}{Output}
  -\mmaFrac{3\,\mmaSup{\(\mu\)}{4}}{2\,lam}\,+\,\mmaFrac{\mmaSup{\(\mu\)}{2}\,\mmaSup{phi}{2}}{2}\,- \mmaFrac{lam \mmaSup{phi}{4}}{24} -\,phi\,\mmaSub{\mmaOver{psi}{_}}{i1,i2}\mmaSub{psi}{i1,i2}
  +\,\mmaFrac{1}{2}\mmaSub{\(\partial\)}{a2}[phi]\mmaSub{\(\partial\)}{mu}[phi]gUp[a2,mu]\,-\,\mmaFrac{1}{4}\mmaSub{\mmaSup{D}{Grav}}{nu}[\mmaSub{A}{a3}]\mmaSub{\mmaSup{D}{Grav}}{a4}[\mmaSub{A}{mu}]
  \,gUp[a3,mu]\,gUp[a4,nu]\,+ ..... -\,e\,\mmaSub{A}{mu}\,\mmaSub{\mmaOver{psi}{_}}{i1,i2}.\mmaSub{psi}{j1,i2}\mmaSup{\mmaSub{\(\gamma\)}{i1,j1}}{v1}
  \,VUp[mu,v1]\,+\,i\,\mmaSub{\mmaOver{psi}{_}}{i1,i2}.\mmaSub{\(\partial\)}{mu}[\mmaSub{psi}{j1,i2}]\,\mmaSup{\mmaSub{\(\gamma\)}{i1,j1}}{v2}\,VUp[mu,v2]\,
  +\,\mmaFrac{1}{8}i\,\mmaSub{\(\partial\)}{mu}[VUp[d1,c1]]\mmaSub{\mmaOver{psi}{_}}{i1,i2}.\mmaSub{psi}{j1,i2}\,VDown[c2,c1]
  \,VUp[mu,v3]\,\mmaSup{\(\gamma\)}{c2}.\mmaSup{\(\gamma\)}{d1}.\mmaSub{\mmaSup{\(\gamma\)}{v3}}{i1,j1} + .....
\end{mmaCell}
}
Herein, \mmaInlineCell{Output}{gUp[a,b]} and \mmaInlineCell{Output}{gDown[a,b]} are upper- and lower-indexed metrics, respectively, \mmaInlineCell{Output}{VUp[a,b]} and \mmaInlineCell{Output}{VDown[a,b]} are upper- and lower-indexed vierbeins, respectively, and \mmaInlineCell{Output}{\mmaSub{\mmaSup{D}{Grav}}{a}[]} is the gravitational covariant derivative. \edit{Notice that we have not included the $\sqrt{-g}$ factor in the Lagrangian; this is because, for simplicity, \code{FeynMG} always assumes this term to be present. Additionally, in this toy model, we have ignored the ghost fields as they decouple in QED; had we included them, they would have been upgraded as shown in \ref{appendix_ghosts}.}

Since the expressions can be long, we will show only the main sections of the output that motivate the next step in the calculation and represent the rest of the terms in ellipsis.

For this example, we will introduce a Brans--Dicke gravitational sector of the form of Eq.~\eqref{ActionGenericJordan}, such that
\begin{equation}\label{eq:action:example}
    \begin{split}
        S=\int \D^4{x} \sqrt{-g} \left[-\frac{\chi}{2}R + \frac{\omega}{2\chi}g^{\mu\nu}\partial_\mu \chi \partial_\nu \chi +\frac{1}{2}\mu_\chi^2 \chi -\frac{\lambda_\chi}{4!}\chi^2 -\frac{3\mu_\chi^4}{2\lambda_\chi}+\code{LCurv}\right],
    \end{split}
\end{equation}
where the $\chi$ field should not be confused with the one defined in Eq.~\eqref{eq_normX}. Before defining the gravitational part of the Lagrangian within \code{FeynMG}, we need to give appropriate attributes to the additional field $\chi$ and the additional parameters (\{$\omega,\mu_\chi,\lambda_\chi$\}). In principle, these can be directly added by updating the model file itself (which should be done before loading it into \code{FeynRules}). Alternatively,  the \code{FeynMG} functions \mmaInlineCell{Input}{\mmaDef{AddScalar[]}} and \mmaInlineCell{Input}{\mmaDef{AddParameter[]}},\footnote{For more information on these functions, see \ref{appendix:functions:output}.} allow the new scalar fields and parameters to be defined after the model file has been loaded into \code{FeynRules}. For the specific case of Eq.~\eqref{eq:action:example}, we need to execute the following:
\begin{mmaCell}[functionlocal=y]{Code}
   \mmaDef{AddScalar}[chi];
   \mmaDef{AddParameter}[muC];
   \mmaDef{AddParameter}[lamC];
   \mmaDef{AddParameter}[w];
\end{mmaCell}
The full Lagrangian can then be defined via
\begin{mmaCell}{Code}
  LJordan= \mmaDef{LCurv} + \mmaDef{chi} \mmaDef{RScalar}/2 + \mmaDef{(w/(2chi)) 
  gUp[Index[Lorentz,mu],Index[Lorentz,nu]]
  del[chi,Index[Lorentz,mu]] del[chi,Index[Lorentz,nu]] 
  + (muC^2chi)/2 - (lamC(chi^2))/(4!)
  - (3muC^4)/(2lamC)};
\end{mmaCell}

In the case of Brans--Dicke-type scalar-tensor theories, it may be convenient to transform to the Einstein frame (see Section \ref{Section_Calculation}). This is achieved in \code{FeynMG} by executing
\begin{mmaCell}{Code}
  LEinstein=\mmaDef{ToEinsteinFrame}[\mmaDef{LJordan}]
\end{mmaCell}
{\footnotesize 
\begin{mmaCell}{Output}
  -\mmaFrac{1}{24}\,lamC\,\mmaSubSup{M}{pl}{4}\,+\,\mmaFrac{\mmaSubSup{M}{pl}{4}\,\mmaSup{muC}{2}}{2\,chi}\,+ ..... -\,\mmaFrac{1}{2}\mmaSubSup{M}{pl}{2}\mmaSub{R}{Sc}\,-\,\mmaFrac{\mmaSubSup{M}{pl}{4}\,phi\,\mmaSub{\mmaOver{psi}{_}}{i1,i2}\mmaSub{psi}{i1,i2}}{\mmaSup{chi}{2}}
  +\,\mmaFrac{3\,\mmaSubSup{M}{pl}{2}\,\mmaSub{\(\partial\)}{a1}[chi]\mmaSub{\(\partial\)}{mu}[chi]gUp[a1,mu]}{4\mmaSup{chi}{2}}\, +\,\mmaFrac{\mmaSubSup{M}{pl}{2}\,w\,\mmaSub{\(\partial\)}{a1}[chi]\mmaSub{\(\partial\)}{mu}[chi]gUp[a1,mu]}{2\mmaSup{chi}{2}}\,
  + ..... -\,\(\alpha\)\,\mmaSub{\mmaSup{D}{Grav}}{a3}[\mmaSub{A}{mu}]\mmaSub{\mmaSup{D}{Grav}}{a4}[\mmaSub{A}{nu}]\,gUp[a3,mu]\,gUp[a4,nu]\,-\,\mmaFrac{e\,\mmaSubSup{M}{pl}{3}\,\mmaSub{A}{mu}}{\mmaSup{chi}{3/2}}
   \,\mmaSub{\mmaOver{psi}{_}}{i1,i2}.\mmaSub{psi}{j1,i2}\mmaSup{\mmaSub{\(\gamma\)}{i1,j1}}{v1}\,VUp[mu,v1]\,+\,\mmaFrac{i\,\mmaSubSup{M}{pl}{3}}{\mmaSup{chi}{3/2}}\,\mmaSub{\mmaOver{psi}{_}}{i1,i2}.\mmaSub{\(\partial\)}{mu}[\mmaSub{psi}{j1,i2}]
   \,\mmaSup{\mmaSub{\(\gamma\)}{i1,j1}}{v2}\,VUp[mu,v2] + ..... +\,\mmaFrac{i\mmaSubSup{M}{pl}{3}}{16\mmaSup{chi}{5/2}}\,\mmaSub{\(\partial\)}{d2}[chi]\,\mmaSub{\mmaOver{psi}{_}}{i1,i2}.\mmaSub{psi}{j1,i2}\,
   \,VDown[c2,mu]\,VUp[d3,d2]\,VUp[mu,v3]\,\mmaSup{\(\gamma\)}{v3}.\mmaSup{\(\gamma\)}{d3}.\mmaSub{\mmaSup{\(\gamma\)}{c2}}{i1,j1}
\end{mmaCell}
}
The output agrees with the result from Eq.~\eqref{ActionEinsteinFull}, including the last term, which comes from the fermion spin-connection [Eq.~\eqref{Ricci_transformations}]. As mentioned before, the Jacobian factor $\sqrt{-g}$ is assumed in the calculation (although it can be omitted by specifying the option \mmaInlineCell{Input}{\{\mmaDef{Jacobian}->\mmaDef{Off}\}} (see \ref{appendix:functions:metrictransf} for further details).  The gravitational sector is now of canonical Einstein--Hilbert form, and we can take the flat spacetime (Minkowski) limit by calling
\begin{mmaCell}{Code}
  \mmaDef{GravityOff}[\mmaDef{LEinstein}]
\end{mmaCell}
{\footnotesize 
\begin{mmaCell}{Output}
  -\mmaFrac{1}{24}\,lamC\,\mmaSubSup{M}{pl}{4}\,+\,\mmaFrac{\mmaSubSup{M}{pl}{4}\,\mmaSup{muC}{2}}{2\,chi}\,+ ..... +\,\mmaFrac{\mmaSubSup{M}{pl}{2}\,w\,\mmaSub{\(\partial\)}{mu}[chi\mmaSup{]}{2}}{2\mmaSup{chi}{2}}\,+\,\mmaFrac{3\,\mmaSubSup{M}{pl}{2}\,\mmaSub{\(\partial\)}{mu}[chi\mmaSup{]}{2}}{4\mmaSup{chi}{2}}\, 
  +\,\mmaFrac{\mmaSubSup{M}{pl}{2}\,\mmaSub{\(\partial\)}{mu}[phi\mmaSup{]}{2}}{2chi}\, +\,\mmaFrac{1}{2}\mmaSub{\(\partial\)}{a3}[\mmaSub{A}{a4}]\mmaSub{\(\partial\)}{a4}[\mmaSub{A}{a3}] - .....
\end{mmaCell}
}
wherein the couplings of the additional scalar field to the matter fields are manifest. The remaining fields are, however, not canonically normalized, and further manipulations are needed in order to pass this Lagrangian back to \code{FeynRules}. These are the focus of the next subsection. 


\subsection{Brans--Dicke theory for \code{FeynRules} in the Jordan frame}
\label{Example_Jordan}

The calculation in the Jordan frame repeats the same steps as in the last subsection up to and including \code{In[4]}:\ We first need to load a model file. We then insert the curvature dependence using \mmaInlineCell{Input}{\mmaDef{InsertCurv}[]} with the Lagrangian as the argument and provide a gravitational sector for the theory. The next step is to expand the metric \edit{around} a flat spacetime background. This can be done by using
\begin{mmaCell}[functionlocal=y]{Code}
  E1=\mmaDef{LinearizeGravity}[\mmaDef{LJordan,{SHGauge->On,UpdDevs->On}}]
\end{mmaCell}
{\small
\begin{mmaCell}{Output}
  -\mmaFrac{\mmaSup{chi}{2}\,lamC}{24}\,+ ..... +\,\mmaFrac{1}{4}\mmaSub{\(\partial\)}{\(\lambda\)1}[chi]\mmaSub{\(\partial\)}{b1}[\mmaSub{h}{b1,\(\lambda\)1}]\,-\,\mmaFrac{1}{8}\,chi\,\mmaSub{\(\partial\)}{mu1}[\mmaSub{h}{b1,\(\lambda\)1}\mmaSup{]}{2}
   -\,\mmaFrac{1}{8}\,chi\,\mmaSub{\(\partial\)}{\(\lambda\)1}[\mmaSub{h}{b1,\(\lambda\)2}\mmaSup{]}{2}+ .....
\end{mmaCell}
}
where \code{LJordan} was defined previously in \code{In[4]}, and the provided options specify that the scalar-harmonic gauge from Eq.~\eqref{GFterm} is used and all covariant derivatives are updated to the modified form from Eq.~\eqref{eq_ModifiedCovDev}. As mentioned previously, the Jacobian $\sqrt{-g}$ has been included when linearizing gravity by default, but it can be omitted using \mmaInlineCell{Input}{\{\mmaDef{Jacobian}->\mmaDef{Off}\}} (see \ref{appendix:functions:perturb}).

As we can see in the second line, many of the terms are repeated, since \code{Mathematica} does not use Einstein's index notation, for which two repeated indices are summed over. As a result, various terms in the output will be equivalent, differing only in their index labels (e.g., $A_\mu A^\mu=A_\rho A^\rho$). In order to force \code{Mathematica} to combine these terms, we have to use the same set of indices for all the terms. This problem is solved by the function \mmaInlineCell{Input}{\mmaDef{IndexSimplify}}:
\begin{mmaCell}[functionlocal=y]{Code} 
  E2=\mmaDef{IndexSimplify}[\mmaDef{E1},{mu,nu,rho}]
\end{mmaCell}
{\small
\begin{mmaCell}{Output}
  ..... -\(\alpha\)\,\mmaSub{A}{mu}\mmaSub{A}{nu}\,CMod[chi\mmaSub{\mmaSup{]}{\{1\}mu}}{rho,rho}\,CMod[chi\mmaSub{\mmaSup{]}{\{1\}nu}}{MG1,MG1}\,+\,\mmaFrac{\mmaSub{\(\partial\)}{mu}[chi\mmaSup{]}{2}}{4\,chi}\,
  +\,\mmaFrac{w\,\mmaSub{\(\partial\)}{mu}[chi\mmaSup{]}{2}}{2\,chi}\,+\,\mmaFrac{1}{2}\mmaSub{\(\partial\)}{mu}[phi\mmaSup{]}{2}\,-\,\mmaFrac{1}{2}\mmaSub{\(\partial\)}{nu}[\mmaSub{A}{mu}\mmaSup{]}{2}\,+ ..... +\,\mmaFrac{1}{8}chi\,\mmaSub{\(\partial\)}{rho}[\mmaSub{h}{mu,nu}\mmaSup{]}{2}
  + ..... -\,\mmaFrac{1}{4}\,\mmaSub{\(\partial\)}{mu}[chi]\,\mmaSub{\(\partial\)}{mu}[h]\,+ ..... -\,\mmaFrac{1}{16}chi\,\mmaSub{\(\partial\)}{mu}[h\mmaSup{]}{2}+ .....
\end{mmaCell}
}
The optional argument \mmaInlineCell{Input}{\mmaDef{\{mu,nu,rho\}}} allows the user to choose a set of $n$ indices from which the first $n$ replacements will be taken.

The output of \mmaInlineCell{Input}{\mmaDef{E2}} contains significantly fewer terms than \mmaInlineCell{Input}{\mmaDef{E1}}. Moreover, \mmaInlineCell{Input}{\mmaDef{E2}} already contains the expected graviton kinetic energy and its kinetic mixing with the scalar field \code{chi}, as in Eq.~\eqref{eq_Jrd_notcanonX}, thanks to the specification of the option \mmaInlineCell{Input}{\{\mmaDef{SHGauge->0n}\}} in \mmaInlineCell{Input}{\mmaDef{LinearizeGravity}} that implements the scalar-harmonic gauge and associated covariant derivatives from Eq.~\eqref{CmodImpl}, which are convenient for the case of pure Brans--Dicke-type theories. This choice has led to the \mmaInlineCell{Input}{\mmaDef{CMod}[]} terms in the Lagrangian, which need to be expanded in terms of the \code{chi} field. However, as described earlier, the truncation to first order in \code{chi} does not commute with non-linear field redefinitions, so the \mmaInlineCell{Input}{\mmaDef{CMod}[]} term will only be expanded once all the fields have their canonical kinetic energy.

We can check that the kinetic energies appearing in \mmaInlineCell{Input}{\mmaDef{E2}} are not canonically normalized by running
\begin{mmaCell}[functionlocal=y]{Code} 
  \mmaDef{CheckCanonScalar}[\mmaDef{E2}]
\end{mmaCell}
{\small
\begin{mmaCell}{Output}
  \mmaFrac{(1+2\,w)\,\,\mmaSub{\(\partial\)}{mu}[chi\mmaSup{]}{2}}{4\,\,chi} + \mmaFrac{1}{2}\,\mmaSub{\(\partial\)}{mu}[phi\mmaSup{]}{2}
  
  There are one or more non-canonical kinetic energies.
  Use CanonScalar.
\end{mmaCell}
}
As the output indicates, we can execute
\begin{mmaCell}[functionlocal=y]{Code} 
  E3=\mmaDef{CanonScalar}[\mmaDef{E2}]
\end{mmaCell}
{\small
\begin{mmaCell}{Output}
  ..... +\,\mmaFrac{\mmaSup{chi}{4}\,lamC}{96(1+2w\mmaSup{)}{2}}\,+\,\mmaFrac{\mmaSup{chi}{2}\,\mmaSup{muC}{2}}{4(1+2w\mmaSup{)}{2}}\,+\,.....\,+\,\mmaFrac{\mmaSup{chi}{2}\,\mmaSup{muC}{2}w}{2(1+2w\mmaSup{)}{2}}\,+\,.....\,+\,\mmaFrac{1}{2}\mmaSub{\(\partial\)}{mu}[chi\mmaSup{]}{2}\,
  -\,\mmaFrac{4\(\alpha\)\mmaSub{A}{mu}\mmaSub{A}{nu}\,\mmaSub{\(\partial\)}{mu}[chi]\mmaSub{\(\partial\)}{nu}[chi]}{\mmaSubSup{v}{chi}{2}}\,+\,\mmaFrac{1}{2}\mmaSub{\(\partial\)}{mu}[phi\mmaSup{]}{2}\,+ ..... +\,\mmaFrac{\mmaSup{chi}{2}\,\mmaSub{\(\partial\)}{rho}[\mmaSub{h}{mu,nu}\mmaSup{]}{2}}{16(1+2\,w)}\,
  +\,.....\,-\,\mmaFrac{chi\,\mmaSub{\(\partial\)}{mu}[chi]\,\mmaSub{\(\partial\)}{mu}[h]}{4(1+2w)}\, + .....-\,\mmaFrac{\mmaSup{chi}{2}\,\mmaSub{\(\partial\)}{mu}[h\mmaSup{]}{2}}{32(1+2w\mmaSup{)}{2}}-\,\mmaFrac{\mmaSup{chi}{2}w\,\mmaSub{\(\partial\)}{mu}[h\mmaSup{]}{2}}{16(1+2w\mmaSup{)}{2}}
  + .....
\end{mmaCell}
}
The kinetic energies of the scalar fields are now canonically normalized, leading to the expansion of every \mmaInlineCell{Input}{\mmaDef{CMod}[]} (where present). This expansion is performed in terms of the scalar field \code{chi}.

At this stage, the kinetic energy of the graviton is composed of multiple terms. These could be simplified by means of \code{Mathematica}'s \mmaInlineCell{Code}{\mmaDef{FullSimplify}} command, but this will often prove time-consuming, and it is not necessary, except for aesthetic reasons. From here, the only thing left to do is to canonically normalize the graviton kinetic energy. To this end, we need to shift the fields around their vevs, so the graviton kinetic energy acquires a constant prefactor. This can be achieved by running
\begin{mmaCell}[functionlocal=y]{Code} 
  E4=\mmaDef{VevExpand}[\mmaDef{E3}]
\end{mmaCell}
{\footnotesize
\begin{mmaCell}[label={}]{Output}
  \(\bigg\{\)1\,==\,\(\big\{\)\mmaSub{v}{chi}->\mmaSub{v}{chi},\mmaSub{v}{phi}->\mmaSub{v}{phi}\(\big\}\)\,,\,2\,==\,\(\big\{\)\mmaSub{v}{chi}->-\mmaFrac{2 \mmaRadical{3}\,\,\mmaRadical{\mmaSup{muC}{2}\,+\,2\,\mmaSup{muC}{2}w}}{\mmaRadical{lamC}},\mmaSub{v}{phi}->0\(\big\}\),
  \,.....\,, \,7\,==\,\(\big\{\)\mmaSub{v}{chi}->-\mmaFrac{2 \mmaRadical{3}\,\,\mmaRadical{\mmaSup{muC}{2}\,+\,2\,\mmaSup{muC}{2}w}}{\mmaRadical{lamC}},\mmaSub{v}{phi}->\mmaFrac{\mmaRadical{6}\,\(\mu\)}{\mmaRadical{lam}}\(\big\}\)\,,\,.....\(\bigg\}\) 
\end{mmaCell}
}
{\small
\begin{mmaCell}{Output}
   -\mmaSup{\(\mu\)}{2}\mmaSup{phi}{2}\,-\,\mmaFrac{\mmaRadical{lam}\,\(\mu\)\,\mmaSup{phi}{3}}{\mmaRadical{6}} + ..... -\,\mmaFrac{\mmaSup{chi}{3}\,\mmaRadical{lamC}\,muC\,}{4 \mmaRadical{3}\,(1+2w\mmaSup{)}{3/2}}\,+\,.....\,
   +\,\mmaFrac{1}{2}\mmaSub{\(\partial\)}{mu}[chi\mmaSup{]}{2}\,-\,\mmaFrac{\(\alpha\)\,lamC\,\mmaSub{A}{mu}\mmaSub{A}{nu}\,\mmaSub{\(\partial\)}{mu}[chi]\mmaSub{\(\partial\)}{nu}[chi]}{3\,\mmaSup{muC}{2}(1+\,2w)}\,+\,\mmaFrac{1}{2}\mmaSub{\(\partial\)}{mu}[phi\mmaSup{]}{2}\,
   + ..... +\,\mmaFrac{3\,\mmaSup{muC}{2}\mmaSub{\(\partial\)}{rho}[\mmaSub{h}{mu,nu}\mmaSup{]}{2}}{4\,lamC}\,+ ..... +\,\mmaFrac{\mmaRadical{3}\,muC\,\mmaSub{\(\partial\)}{mu}[chi]\mmaSub{\(\partial\)}{mu}[h]}{2\,\mmaRadical{lamC}\,\,\mmaRadical{1+2\,w}}\,
   + .....
\end{mmaCell}
}
Note that this function shows all the extrema of the potential. Since there may be multiple minima, the function allows the user to choose which vev (or set of vevs) will be used by a dialogue window prompt (in this case, we choose option 7). Notice that the \mmaInlineCell{Input}{\mmaSub{\mmaDef{v}}{\mmaDef{chi}}} dependence already present from the expansion of the \code{CMod} functions has also been replaced by the user-selected vev in \mmaInlineCell{Input}{\mmaDef{E3}}.

Once we have a constant prefactor to the graviton kinetic energy, we can canonically normalize it using
\begin{mmaCell}[functionlocal=y]{Input} 
  E5=\mmaDef{CanonGrav}[\mmaDef{E4}]
\end{mmaCell}
{\small
\begin{mmaCell}{Output}
    -\mmaSup{\(\mu\)}{2}\mmaSup{phi}{2}\,-\,\mmaFrac{\mmaRadical{lam}\,\(\mu\)\,\mmaSup{phi}{3}}{\mmaRadical{6}}\,+ ..... -\,\mmaFrac{\mmaSup{chi}{3}\,\mmaRadical{lamC}\,muC\,}{4 \mmaRadical{3}\,(1+2w\mmaSup{)}{3/2}}\,+\,.....\,+\,\mmaFrac{1}{2}\mmaSub{\(\partial\)}{mu}[chi\mmaSup{]}{2}\,
    -\,\mmaFrac{\(\alpha\)\,lamC\,\mmaSub{A}{mu}\mmaSub{A}{nu}\,\mmaSub{\(\partial\)}{mu}[chi]\mmaSub{\(\partial\)}{nu}[chi]}{3\,\mmaSup{muC}{2}(1+\,2w)}\,+\,\mmaFrac{1}{2}\mmaSub{\(\partial\)}{mu}[phi\mmaSup{]}{2}\,+\,.....\,+\,\mmaFrac{1}{2}\mmaSub{\(\partial\)}{rho}[\mmaSub{h}{mu,nu}\mmaSup{]}{2}\,
    + ..... -\,\mmaFrac{\mmaSub{\(\partial\)}{mu}[chi]\mmaSub{\(\partial\)}{mu}[h]}{\mmaRadical{2}\,\mmaRadical{1+2\,w}}\,+ ..... -\,\mmaFrac{1}{4}\mmaSub{\(\partial\)}{mu}[h\mmaSup{]}{2}\,+  .....
\end{mmaCell}
}
We have recovered the usual canonically normalized Fierz--Pauli kinetic energy terms from Eq.~\eqref{Eq_FiewzP}. We also see the expected kinetic mixing between the scalar field and the graviton, which can be identified by executing
\begin{mmaCell}[functionlocal=y]{Code} 
  \mmaDef{CheckGravityMixing}[\mmaDef{E5}]
\end{mmaCell}
{\small
\begin{mmaCell}{Output}
  -\mmaFrac{\mmaSub{\(\partial\)}{mu}[chi]\mmaSub{\(\partial\)}{mu}[h]}{\mmaRadical{2\,+\,4\,w}} 
  
  There are kinetic mixing terms for gravity. 
  Use GravKinMixing.
\end{mmaCell}
}

The final manipulation is to diagonalize this kinetic mixing. This can be achieved by running (see \ref{appendix_C} for a detailed description of how this operation is done) 
\begin{mmaCell}[functionlocal=y]{Code} 
  E6=\mmaDef{GravKinMixing}[\mmaDef{E5},{\mmaDef{OutSimplify->On}}]
\end{mmaCell}
{\small
\begin{mmaCell}{Output}
    -\mmaSup{\(\mu\)}{2}\mmaSup{phi}{2}\,-\,\mmaFrac{\mmaRadical{lam}\,\(\mu\)\,\mmaSup{phi}{3}}{\mmaRadical{6}}\,+ ..... +\,\mmaFrac{1}{2}\mmaSub{\(\partial\)}{mu}[chi\mmaSup{]}{2}\,+ ..... 
    +\,\mmaFrac{1}{2}\mmaSub{\(\partial\)}{mu}[phi\mmaSup{]}{2}\,-\,\mmaFrac{chi\,\mmaRadical{lamC}\,\mmaSub{\(\partial\)}{mu}[phi\mmaSup{]}{2}}{2\,\mmaRadical{3}\,muC\,\,\mmaRadical{3\,+\,2w}}\,+.....+\,\mmaFrac{1}{2}\mmaSub{\(\partial\)}{rho}[\mmaSub{h}{mu,nu}\mmaSup{]}{2}\,
    + ..... -\,\mmaFrac{1}{4}\mmaSub{\(\partial\)}{mu}[h\mmaSup{]}{2}\,+..... -\,\mmaFrac{\mmaRadical{6}\,\(\mu\)\,\mmaSub{\mmaOver{psi}{_}}{i1,i2}.\mmaSub{psi}{i1,i2}}{\mmaRadical{lam}}\,
    +\,.....\,+\,\mmaFrac{2\,\,\mmaRadical{2}\,chi\,\mmaRadical{lamC}\,\(\mu\)\,\mmaSub{\mmaOver{psi}{_}}{i1,i2}.\mmaSub{psi}{i1,i2}}{\mmaRadical{lam}\,muC\,\,\mmaRadical{3\,+\,2w}}\,- .....
    -\mmaFrac{i\,\mmaRadical{3}\,chi\,\mmaRadical{lamC}\,\mmaSub{\mmaOver{psi}{_}}{i1,i2}.\mmaSub{\(\partial\)}{mu}[\mmaSub{psi}{j1,i2}] \mmaSubSup{\(\gamma\)}{i1,j1}{mu}}{2\,muC\,\,\mmaRadical{3\,+\,2\,w}}\, + .....
\end{mmaCell}
}
The argument \mmaInlineCell{Input}{\{{\mmaDef{OutSimplify->On}}\}} applies \mmaInlineCell{Input}{\mmaDef{FullSimplify[]}} up to quadratic terms, so that the kinetic energy terms appear explicitly canonicalized. Note that this simplification is not a prerequisite to further processing of the output with \code{FeynRules}.

\code{FeynMG} can extract the effective $\Mpl$ at any point in the calculations (before or after linearizing gravity or canonically normalizing the kinetic energies). For example, for the diagonalized Lagrangian from \code{Out[15]} (corresponding to \mmaInlineCell{Input}{\mmaDef{E6}}), we obtain
\begin{mmaCell}[functionlocal=y]{Code} 
  \mmaDef{GiveMpl}[\mmaDef{E6}]
\end{mmaCell}
{\small
\begin{mmaCell}{Output}
  Using the values defined in the parameter and mass classes, the
  effective value for Mpl in this Lagrangian is Sqrt[6] GeV.
  It can be changed by modifying the Model File obtained after
  using OutputModelMG.

  -\mmaFrac{\mmaRadical{6}\,muC}{\mmaRadical{lamC}}
\end{mmaCell}
}
The effective value for the Planck mass in the inputted Lagrangian is automatically calculated and printed using the defined values of the loaded parameters. In this case, since both parameters \{\mmaInlineCell{Output}{\mmaDef{muC,lamC}}\} were defined using \mmaInlineCell{Input}{\mmaDef{AddParameter[]}}, their values were set by default to 1.\footnote{When using \code{AddParameter[]}, the value of the parameter can be specified using the option \code{Value$\to$X} For more information, see \ref{appendix:functions:output}}. Moreover, we can substitute the calculated value for $\Mpl$ into the Lagrangian by calling
\begin{mmaCell}[functionlocal=y]{Code} 
  \mmaDef{InsertMpl}[\mmaDef{E6}]
\end{mmaCell}
{\small
\begin{mmaCell}{Output}
  Using the values defined in the parameter and mass classes, the
  effective value for Mpl in this Lagrangian is Sqrt[6] GeV.
  It can be changed by modifying the Model File obtained after
  using OutputModelMG.
  
  -\mmaSup{\(\mu\)}{2}\mmaSup{phi}{2}\,-\,\mmaFrac{\mmaRadical{lam}\,\(\mu\)\,\mmaSup{phi}{3}}{\mmaRadical{6}}\,+ ..... +\,\mmaFrac{1}{2}\mmaSub{\(\partial\)}{mu}[chi\mmaSup{]}{2}\,+ ..... 
    +\,\mmaFrac{1}{2}\mmaSub{\(\partial\)}{mu}[phi\mmaSup{]}{2}\,-\,\mmaFrac{chi\,\mmaSub{\(\partial\)}{mu}[phi\mmaSup{]}{2}}{\mmaRadical{2}\,\mmaSub{M}{pl}\,\,\mmaRadical{3\,+\,2w}}\,+\,.....\,+\,\mmaFrac{1}{2}\mmaSub{\(\partial\)}{rho}[\mmaSub{h}{mu,nu}\mmaSup{]}{2}\,
    + ..... -\,\mmaFrac{1}{4}\mmaSub{\(\partial\)}{mu}[h\mmaSup{]}{2}\,+ ..... -\,\mmaFrac{\mmaRadical{6}\,\(\mu\)\,\mmaSub{\mmaOver{psi}{_}}{i1,i2}.\mmaSub{psi}{i1,i2}}{\mmaRadical{lam}}\,
    + ..... +\,\mmaFrac{4\,\,\mmaRadical{3}\,chi\,\(\mu\)\,\mmaSub{\mmaOver{psi}{_}}{i1,i2}.\mmaSub{psi}{i1,i2}}{\mmaRadical{lam}\,\mmaSub{M}{pl}\,\,\mmaRadical{3\,+\,2w}}\,
    - .....  +\,i\,\mmaSub{\mmaOver{psi}{_}}{i1,i2}.\mmaSub{\(\partial\)}{mu}[\mmaSub{psi}{j1,i2}] \mmaSubSup{\(\gamma\)}{i1,j1}{mu}
    -\mmaFrac{3\,i\,chi\,\mmaSub{\mmaOver{psi}{_}}{i1,i2}.\mmaSub{\(\partial\)}{mu}[\mmaSub{psi}{j1,i2}] \mmaSubSup{\(\gamma\)}{i1,j1}{mu}}{\mmaRadical{2}\,\mmaSub{M}{pl}\,\,\mmaRadical{3\,+\,2w}}\,- .....
\end{mmaCell}
}
This function also automatically adds $\Mpl$ to the list of defined parameters, with its corresponding value ($\sqrt{6}$GeV for this specific case) and \code{InteractionOrder$\to$-1}. Notice that a Yukawa coupling between the fermion fields and the \code{chi} field has appeared in the fourth line, as expected. However, a closer look at this term shows that the coupling constant is four times larger than the result $m_\psi/\sqrt{2\Mpl^2(2+3w)}$ from Refs.~\cite{Burrage:2018dvt,Copeland:2021qby}. This is because of the last term in the expression, which will also contribute to the tree-level interactions between the fermion and the scalar field, leading then to the same results as in Refs.~\cite{Burrage:2018dvt,Copeland:2021qby}.

At this point, all the interactions up to second order in the fields have been canonically normalized and diagonalized, so there are no kinetic or mass mixings. Therefore, the updated Lagrangian for the matter fields with the additional scalar field couplings is now in a form that can be processed further by \code{FeynRules} and compatible packages for phenomenological studies.


\subsection{Outputting a model file}

\code{FeynMG} allows the user to create a new model file with the Lagrangian of their choice, in which all the introduced particles (such as the graviton and additional scalar field) and new parameters (such as $\Mpl$) will be incorporated and properly defined.\footnote{New particles and parameters created using \code{AddScalar[]} and \code{AddParameter[]} will also be added, see \ref{appendix:functions:output} for more information.} This can be done by running
\begin{mmaCell}[functionlocal=y]{Code} 
  \mmaDef{OutputModelMG}[OldModelFile,NewModelFile,Lagrangian];
\end{mmaCell}
where \mmaInlineCell{Input}{OldModelFile} is the name of the original \code{FeynRules} model file that the user loaded, \mmaInlineCell{Input}{NewModelFile} is the chosen name of the new model file, and \mmaInlineCell{Input}{Lagrangian} is the final Lagrangian, as prepared with \code{FeynMG}.

The upgraded model file can be read directly into \code{FeynRules} without needing to load or rerun \code{FeynMG}.


\section{Conclusions}
\label{sec:conclusions}

Modifying the gravitational sector of a Lagrangian can lead to new interactions between matter fields that need not be Planck-suppressed, but making these interactions manifest by hand on a model-by-model basis is tedious and time-consuming. In this paper, we have presented the \code{Mathematica} package \code{FeynMG}, which can manipulate scalar-tensor theories of gravity into a format that can be processed by \code{FeynRules}.

Even for the simplest toy models, it is necessary to perform transformations of the metric or linearize the gravitational action, redefine multiple fields, expand around the vacuum expectation values of the scalar fields and diagonalize mass and/or kinetic mixings, in particular those between additional scalar fields and the trace of the graviton. \code{FeynMG} provides a set of functions that allow the user to recycle existing \code{FeynRules} model files that do not contain gravity and to implement these various steps.

Once the user arrives at a canonically normalized Lagrangian, in which all kinetic and mass mixings have been diagonalized, it can be further processed by \code{FeynRules} and compatible packages to allow phenomenological studies of scalar-tensor theories of gravity. Moreover, instead of deriving the same Lagrangian every time one uses \code{Mathematica}, \code{FeynMG} allows the output of a new model file with all the updated fields, parameters and chosen Lagrangian. A summary list of functions can be found in \ref{appendix:functions}.

In this paper, we have described the implementation of a minimal example in \code{FeynMG}: Brans--Dicke theory coupled to  QED plus a real scalar prototype of the Standard Model Higgs. The inbuilt functions, however, may be used to manipulate more complicated gravitational sectors, such as multi-field scalar-tensor theories or Horndeski theories, and additional functionality is being developed for future release.


\section*{Acknowledgments}

The authors thank Clare Burrage, Christoph Englert, Oliver Gould, Phil Parry and Yannick Kluth for helpful discussions. This work was supported by an IPPP Associateship from the Institute for Particle Physics Phenomenology (IPPP) at Durham University, a Nottingham Research Fellowship from the University of Nottingham, a United Kingdom Research and Innovation (UKRI) Future Leaders Fellowship [Grant Nos.~MR/V021974/1 and MR/V021974/2],  Science and Technology Facilities Council (STFC) Consolidated Grants [Grant Nos.~ST/T000732/1 and~ST/X00077X/1], a Leverhulme Research Fellowship [RF-2021 312], and an STFC studentship [Grant No.~ST/V506928/1]. For the purpose of open access, the authors have applied a Creative Commons Attribution (CC BY) license to any Author Accepted Manuscript version arising.


\section*{Data availability}

The package \code{FeynMG} is available at \href{https://gitlab.com/feynmg/FeynMG}{gitlab.com/feynmg/FeynMG}; it makes use of the existing package \code{FeynRules}, which is openly available from \href{https://feynrules.irmp.ucl.ac.be}{feynrules.irmp.ucl.ac.be}.
	

\begin{appendix}
\section{Modified covariant derivative and U(1) Feynman gauge}\label{appendix_A}

In Section \ref{Subsection Calculation Jordan}, we described an update to the covariant derivative in the Jordan frame, based on Ref.~\cite{Copeland:2021qby, DeWitt:1964mxt, Barvinsky:1985an, Finn:2022rlo}, that proves convenient for Brans--Dicke-type theories with only a non-minimal coupling to the Ricci scalar. This modified covariant derivative $\mathcal{D}_{\mu}$ reduces to the usual $\nabla_\mu$ when Weyl transformed to the Einstein frame. This modified covariant derivative is given by
\begin{equation}\label{eq:appendix:modcovdev}
    \mathcal{D}_\mu Y_\nu= \nabla_\mu Y_\nu \edit{-} C^\rho_{\mu\nu} Y_\rho,
\end{equation}
where 
	\begin{equation}
	    C^\rho_{\mu\nu}=\frac{1}{2X}(\delta^\rho_\mu\partial_\nu X+\delta^\rho_\nu\partial_\mu X - g_{\mu\nu}\partial^{\rho} X),
	\end{equation}
	and it allows us to define the so-called scalar-harmonic gauge, which maps to the usual harmonic gauge in the Einstein frame.

 We can proceed similarly with other gauge fixing terms, such as the one for the U(1) gauge field. For instance, defined in the Einstein frame where gravity is canonical, the Feynman gauge fixing action takes the following form:
\begin{equation}
    S\supset\int d^4{x} \sqrt{-\tilde{g}}\,\,\tilde{g}^{\mu\nu}\tilde{g}^{\sigma\rho}\tilde{\nabla}_\mu A_\nu\tilde{\nabla}_\sigma A_\rho,
\end{equation}
where all the tilded objects are built with the Einstein frame metric $\tilde{g}_{\mu\nu}$. On transforming to the Jordan frame, we would find that the gauge fixing term has to be written as
\begin{equation}
    S\supset\int d^4{x} \sqrt{-{g}}\,\,{g}^{\mu\nu}{g}^{\sigma\rho}{\mathcal{D}}_\mu A_\nu{\mathcal{D}}_\sigma A_\rho,
\end{equation}
where the covariant derivatives have transformed to their modified forms from \eqref{eq:appendix:modcovdev}, which will introduce new couplings between the gauge field $A_\mu$ and the scalar field that appears in the Weyl rescaling of the metric. These new couplings are encoded in the $C^\rho_{\mu\nu}$ terms. In what follows, we will show that these new interactions are those that ensure there are no interactions between the scalar field and the gauge field at dimension four, as we would expect from general arguments based on Weyl invariance.

    Let us first define the object $\Delta_{\lambda\mu\nu}$ via
    \begin{equation}
     \frac{1}{2}g^{\rho\lambda}\Delta_{\lambda\mu\nu}=\Gamma^\rho_{\mu\nu}+C^\rho_{\mu\nu},
    \end{equation}
    where we have taken the common factor of the upper-indexed metric so the result can be generalized to any order of its expansion in $h^{\mu\nu}$. For the case of Brans--Dicke theory with a coupling function $F(X)$, we find
    \begin{equation}
        \Delta_{\lambda\mu\nu}=\partial_\mu g_{\lambda\nu}+\partial_\nu g_{\mu\lambda}-\partial_\lambda g_{\mu\nu} +\frac{F'(X)}{F(X)}(g_{\lambda\nu}\partial_\mu X +g_{\mu\lambda}\partial_\nu X-g_{\mu\nu}\partial_\lambda X)
        \label{Delta_tensor}
    \end{equation}
    reduces to
     \begin{equation}
        \Delta_{\lambda\mu\nu}=\partial_\mu h_{\lambda\nu}+\partial_\nu h_{\mu\lambda}-\partial_\lambda h_{\mu\nu} +\frac{F'(X)}{F(X)}(\eta_{\lambda\nu}\partial_\mu X +\eta_{\mu\lambda}\partial_\nu X-\eta_{\mu\nu}\partial_\lambda X)
    \end{equation}
     on perturbing the metric around a flat background (Eq.~\eqref{eq:expansion}). After canonically normalizing the $X$ field through the redefinition from  Eq.~\eqref{eq_normX}, i.e.,
     \begin{equation}
	    \chi(X)=\int_{0}^{X}\D\hat{X}\sqrt{Z(\hat{X})+\frac{F'(\hat{X})^2}{2F(\hat{X})}},
	\end{equation}
    this leads to
         \begin{equation}
        \Delta_{\lambda\mu\nu}=\partial_\mu h_{\lambda\nu}+\partial_\nu h_{\mu\lambda}-\partial_\lambda h_{\mu\nu} +\frac{\hat{F}'(\chi)}{\hat{F}(\chi)}(\eta_{\lambda\nu}\partial_\mu \chi +\eta_{\mu\lambda}\partial_\nu \chi-\eta_{\mu\nu}\partial_\lambda \chi)
        ,\label{Delta_tensor_canon}
    \end{equation}
    where we have defined $\hat{F}(\chi)\equiv F(X)$ and $\hat{F}'(\chi)\equiv \partial\hat{F}(\chi)/\partial \chi$. As described in Section \ref{Subsection Calculation Jordan}, we now expand the scalar field around its vev, so that the graviton can also be canonically normalized [see Eq.~\eqref{eq_Jordan_mixing}, where the kinetic mixing between the graviton and $\chi$ is manifest]. At this point, $\Delta_{\lambda\mu\nu}$ has the form
     \begin{align}
        \Delta_{\lambda\mu\nu}=&\partial_\mu h_{\lambda\nu}+\partial_\nu h_{\mu\lambda}-\partial_\lambda h_{\mu\nu}\nonumber\\ +&\frac{\hat{F}'(v_\chi)}{\hat{F}(v_\chi)+\chi\hat{F}'(v_\chi)}(\eta_{\lambda\nu}\partial_\mu \chi +\eta_{\mu\lambda}\partial_\nu \chi-\eta_{\mu\nu}\partial_\lambda \chi)
        .
        \end{align}

    The kinetic mixing between the graviton and the scalar can be removed [see Eq.\eqref{eq_diagonalization_h_chi}] by means of the transformations in Eq.~\eqref{eq_diagonalization_h_chi}. With this, we obtain Eq.~\eqref{eq_Jordan_fin} and
    \begin{align}
         \label{Delta_tensor_diag}
        \Delta_{\lambda\mu\nu}=&\frac{2}{M_{\rm{Pl}}}\bigg(\partial_\mu h_{\lambda\nu}+\partial_\nu h_{\mu\lambda}-\partial_\lambda h_{\mu\nu}\bigg)\nonumber\\
        +&\frac{1}{\Mpl}\frac{\hat{F}'(v_\chi)}{\sqrt{\Mpl^2+\hat{F}'(v_\chi)^2}}\left(\eta_{\lambda\nu}\partial_\mu \sigma +\eta_{\mu\lambda}\partial_\nu \sigma-\eta_{\mu\nu}\partial_\lambda \sigma\right)\nonumber\\
        -&\frac{\hat{F}'(v_\chi)}{\hat{F}'(v_\chi)\sigma+\Mpl\sqrt{\Mpl^2+\hat{F}'(v_\chi)^2}}(\eta_{\lambda\nu}\partial_\mu \sigma +\eta_{\mu\lambda}\partial_\nu \sigma-\eta_{\mu\nu}\partial_\lambda \sigma)
        ,
    \end{align}
    where $\hat{F}(v_\chi)= \Mpl^2$ has been substituted and $\sigma$ corresponds to the canonically normalized additional scalar field. We can now expand the denominator in the third line up to first order in $\Mpl^{-1}$ to give
    \begin{align}
        \Delta_{\lambda\mu\nu}=&\frac{2}{M_{\rm{Pl}}}\left(\partial_\mu h_{\lambda\nu}+\partial_\nu h_{\mu\lambda}-\partial_\lambda h_{\mu\nu}\right),
    \end{align}
    showing a perfect cancellation of the couplings to the additional scalar. Thus, after diagonalizing, the covariant derivative takes the following form:
    \begin{equation}
        \mathcal{D}_\mu A_\nu=\partial_\mu A_\nu \edit{-}\frac{2}{M_{\rm{Pl}}}\Gamma^\rho _{\mu\nu} A_\rho,
    \end{equation}
    which is nothing but the standard covariant derivative $\nabla_\mu A_\nu$ from Einstein gravity. This is as we would expect, since the diagonalization is essentially a perturbative implementation of the Weyl transformation to the Einstein frame.
    
    We can obtain the same result without diagonalizing and instead summing over all insertions of the graviton-scalar kinetic mixing. Our calculations have shown that the following two series of diagrams cancel with each other:\\
    \begin{equation*}
    \begin{gathered}\vspace{-1mm}
        \includegraphics[scale=0.18]{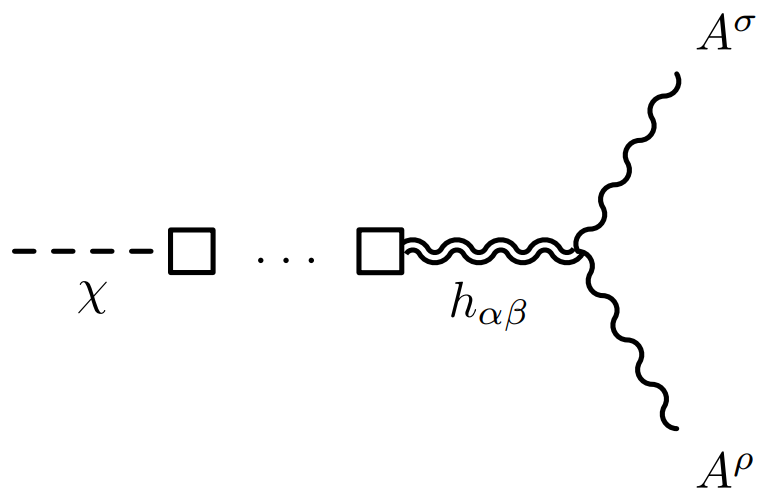}
    \end{gathered}
    +
    \begin{gathered}\vspace{-1.5mm}
        \includegraphics[scale=0.18]{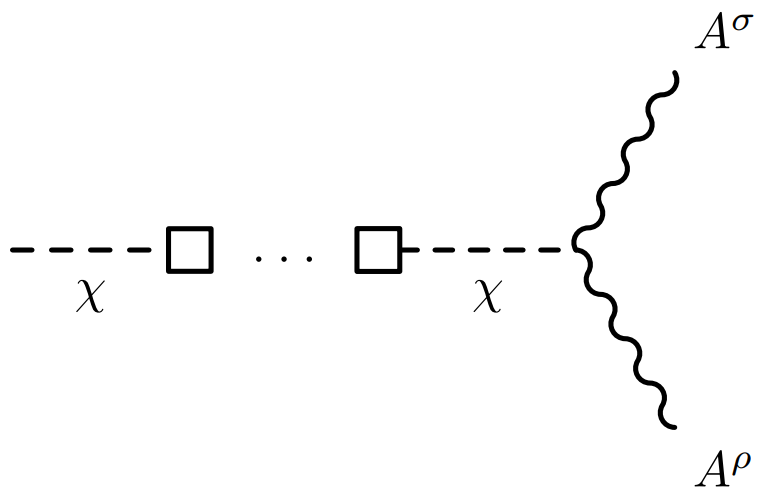}
    \end{gathered}=0,
    \end{equation*}
    where the ellipsis contains the sum over the infinite series of insertions of  mixings (where zero kinetic mixing is also included for the diagram on the right). Similarly, from the diagrams above, we can calculate the incoming graviton amplitude by inserting an additional kinetic mixing to the left of the $\chi$ propagators. Thus, we find that all the diagrams containing kinetic mixings will end up cancelling each other, leaving just the diagram with no kinetic mixings. Diagrammatically, this implies that
    \begin{equation*}
    \begin{gathered}\vspace{-1mm}
        \includegraphics[scale=0.18]{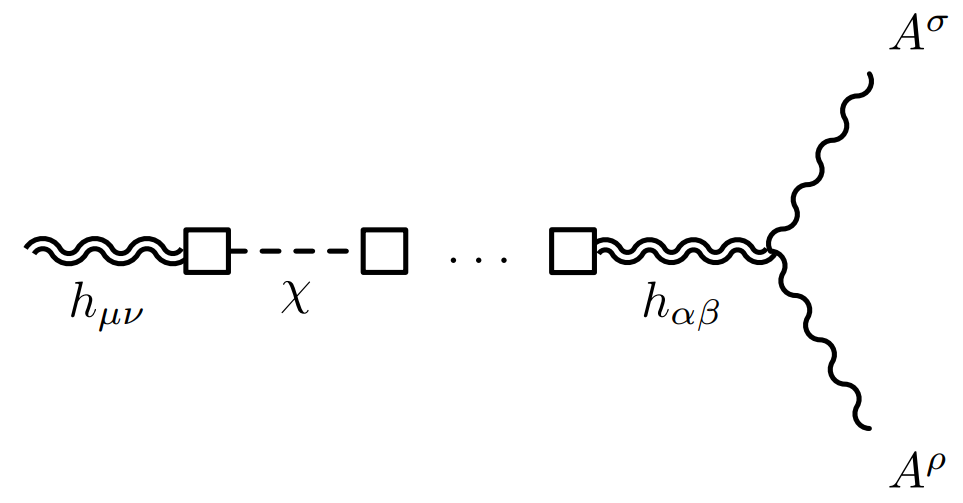}
    \end{gathered}
    +
    \begin{gathered}\vspace{-1mm}
        \includegraphics[scale=0.18]{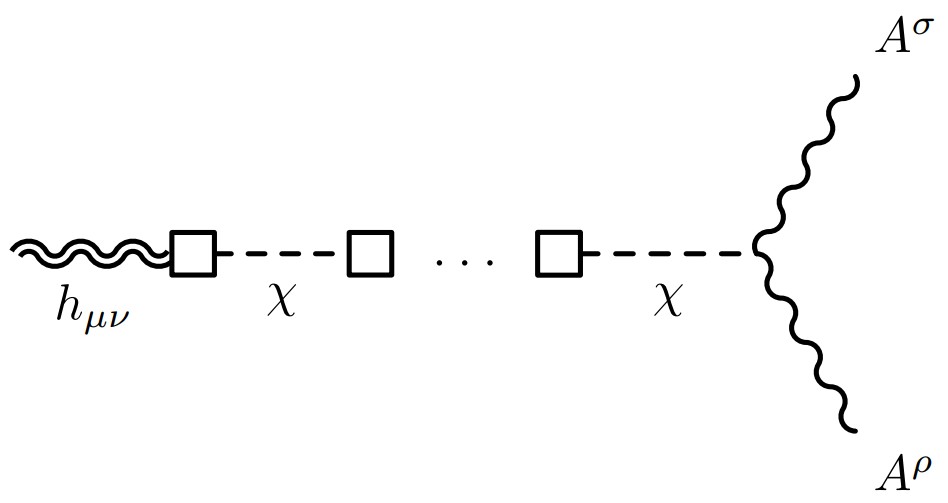}
    \end{gathered}=
    \begin{gathered}\vspace{-3mm}
        \includegraphics[scale=0.18]{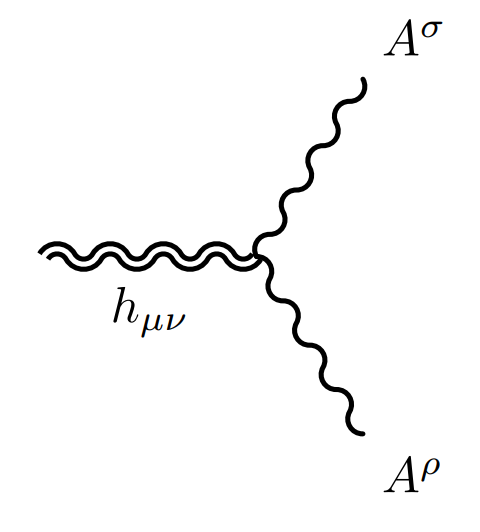}
    \end{gathered},
    \end{equation*}
     which corresponds to the Feynman diagram for the coupling between the gauge field and gravity through the usual Chistoffel symbols.
     
     In either case, we see that the role of the additional terms arising from $C^\rho_{\mu\nu}$ in the updated covariant derivatives is to maintain the Weyl invariance of the Maxwell Lagrangian (at dimension four) once gauge fixing terms are included in the Jordan frame.
    

\section{\edit{Gauge symmetry preservation in scalar-tensor theories}}\label{appendix_ghosts}

With the exception of~\ref{appendix_A}, we have so far ignored any implications that the extension of gravity may have on the existing symmetries of the matter Lagrangian. In this Appendix, we will show that, for the scalar-tensor theories considered in this paper (i.e., those connected to an Einstein frame via a conformal transformation), the gauge symmetries in the matter Lagrangian are not broken due to the modifications of gravity. To do so, we will focus on the Faddeev--Popov procedure, where the introduction of ghost fields eliminates unphysical degrees of freedom~\cite{Freese}.

For generality, we consider the following (non-Abelian) Yang--Mills Lagrangian
\begin{equation}
\lgr_{\rm YM}=-\frac{1}{4}F_{\mu\nu}^a{F^{\mu\nu}}^a-\frac{1}{2\xi}\partial_\mu {A^a}^\mu \partial_\nu {A^a}^\nu -\bar{c}^a\partial^\mu D_\mu c^a,\; 
\end{equation}
where $A^a_\mu$ is the SU$(N)$ gauge field, where the $a$ takes values from $1$ to $N$; $\xi$ is a Lagrange multiplier; $\bar{c}^a$ and $c^a$ are the Faddeev--Popov ghost fields;
\begin{align}
    F_{\mu\nu}^a&=\partial_\mu A_\nu -\partial_\nu A_\mu + g f^{abc}A^b_\mu A^c_\nu,\\
    D_\mu c^a&= \partial_\mu c^a + g f^{abc} A^b_\mu c^c\;,
\end{align}
are the strength tensor and SU$(N)$ covariant derivative (which reduces to Eq.~\eqref{eq:Dpsi} in the U$(1)$ case), respectively; $g$ is a dimensionless constant; and $f^{abc}$ are the SU$(N)$ structure constants. As in Section~\ref{sec:method}, and to study the implications of modified gravity, we first have to insert all minimal couplings, upgrade all partial derivatives to gravitational covariant ones, and append a gravitational sector. For this, we will consider the modified gravitational action from Eq.~\eqref{ActionGenericJordan}, leading to
\begin{align}
    S_{\rm YM}\supset\int d^4{x} \sqrt{-{g}}\,\,&\left[-\frac{F(\chi)}{2}R -\frac{1}{4}g^{\mu\nu}g^{\alpha\beta}F_{\mu\alpha}^a{F_{\nu\beta}^a}\right.\nonumber\\
    &\,\left.-\frac{1}{2\xi}g^{\mu\nu}g^{\alpha\beta}\mathcal{D}_\alpha A^a_\beta \mathcal{D}_\mu A^a_\nu -\bar{c}^ag^{\mu\nu}\mathcal{D}_\mu D_\nu c^a\right]\;,\label{eq:YMJordan}
\end{align}
where we have used the updated covariant derivatives defined in Eq.~\eqref{eq_ModifiedCovDev}. As we will show, this choice for the covariant derivative is not required to preserve the symmetries of the matter Lagrangian. However, we make this choice for convenience, as it eliminates all modifications entering through the Christoffel symbols in the linearized and diagonalized Jordan-frame Lagrangian, as shown in~\ref{appendix_A}. 

Notice that, while the SU$(N)$ gauge symmetry is broken by the gauge-fixing term, the inclusion of the ghost leads to the conservation of an exact global symmetry, called BRST invariance, as a residue of the existing SU$(N)$ symmetry. The symmetry transformations are well-known both for flat spacetimes, and extending them to curved backgrounds is straightforward:\ we just need to insert the minimal couplings to gravity and upgrade the partial derivatives in the transformations, as we did for the action~\cite{Prinz:2020nru}. In this way, we find the following infinitesimal BRST transformations for each field in curved backgrounds:
\begin{subequations}\label{eq:rule1}
\begin{align}
    A^a_\mu &\to A^a_\mu +\frac{1}{g}\theta D_\mu c^a\\
    \bar{c}^a& \to\bar{c}^a-\frac{1}{g\xi}\theta g^{\mu\nu}\mathcal{D}_\mu A^a_\nu\\
    {c}^a&\to{c}^a-\frac{1}{2}\theta f^{abc}c^a c^b.\\
    D_\mu c^a&\to D_\mu c^a\;,
\end{align}
\end{subequations}
where $\theta$ is a Grassmann variable satisfying $\theta^2=0$, and the omitted fields do not transform under this symmetry. Using this set of transformations, we find the following variations of each term:
\begin{subequations}
\begin{align} \delta\left(g^{\mu\nu}g^{\alpha\beta}{F_{\mu\alpha}}^a{F_{\nu\beta}^a}\right)&\to 0\;,\\
    \delta\left(\frac{1}{2\xi}g^{\mu\nu}g^{\alpha\beta}\mathcal{D}_\alpha A^a_\beta \mathcal{D}_\mu A^a_\nu\right)&\to -\frac{\theta}{g\xi}g^{\mu\nu}g^{\alpha\beta}(\mathcal{D}_\mu A^a_\nu)(\mathcal{D}_\alpha D_\beta c^a)\;,\\
    \delta\left(-\bar{c}^ag^{\mu\nu}\mathcal{D}_\mu D_\nu{c}^a\right)&\to \frac{\theta}{g\xi}g^{\mu\nu}g^{\alpha\beta}(\mathcal{D}_\mu A^a_\nu)(\mathcal{D}_\alpha D_\beta c^a)\;.
\end{align}
\end{subequations}
Substituting these results in the action, we can confirm that this set of transformations is indeed a symmetry of the Yang--Mills action, even when gravity is modified as in Eq.~\eqref{eq:YMJordan}. As mentioned, since this derivation did not require the expansion of $\mathcal{D}_\mu$, it is generic for any choice of the gravitational covariant derivative. Moreover, given that the linearization of gravity and the subsequent Lagrangian manipulations shown in Section~\ref{sec:method} do not break any of the SU$(N)$ symmetries, we should expect the BRST symmetry to be maintained at all steps of the calculation, as long as we also perform these operations to the transformations from Eqs.~\eqref{eq:rule1}.

Furthermore, these results can be extended to the Einstein frame. This can be shown by making the Weyl transformation from Eq.~\eqref{eq-metric-transf} to the Jordan-frame action in Eq.~\eqref{eq:YMJordan}. With this, we obtain the following Einstein-frame action:
\begin{align}
    S_{\rm YM}\supset\int d^4{x} \sqrt{-\tilde{g}}\,\,&\left[-\frac{\tMpl^2}{2}\tilde{R} -\frac{1}{4}\tilde{g}^{\mu\nu}\tilde{g}^{\alpha\beta}F_{\mu\alpha}^a{F_{\nu\beta}^a}\right.\nonumber\\
    &\,\left.-\frac{1}{2\xi}\tilde{g}^{\mu\nu}\tilde{g}^{\alpha\beta}\tilde\nabla_\alpha A^a_\beta \tilde\nabla_\mu A^a_\nu -\frac{\tMpl}{F(\chi)}\bar{c}^a\tilde{g}^{\mu\nu}\tilde\nabla_\mu D_\nu c^a\right]\;,\label{eq:YMEins}
\end{align}
where all tilded quantities are built using the Einstein-frame metric $\tilde{g}_{\mu\nu}$, and we can see that the modified covariant derivative $\mathcal{D}_\mu$ has transformed into the usual Christoffel symbols. Note that we work with the Lagrange multiplier field already evaluated on-shell.

While the SU$(N)$ field-strength tensor and gauge-fixing terms do not pick up any $\chi$ couplings, the same is not true for the ghost terms. However, the BRST invariance is still maintained under a modified set of field transformations. In particular, these are
\begin{subequations}\label{eq:rule1Eins}
    \begin{align}
    A^a_\mu\to & A^a_\mu +\frac{1}{g}\theta D_\mu c^a\;,\\
    \bar{c}^a\to&\bar{c}^a-\frac{F(X)}{g\xi\tMpl}\theta \tilde{g}^{\mu\nu}\tilde\nabla_\mu A^a_\nu\;,\\
    {c}^a\to&{c}^a-\frac{1}{2}\theta f^{abc}c^b c^c\;,\\
     D_\mu c^a\to & D_\mu c^a\;,
\end{align}
\end{subequations}
which is nothing but the Weyl rescaling of the Jordan-frame BRST transformations from Eqs.~(\ref{eq:rule1}), as we would expect. In this way, the different terms in Eq.~\eqref{eq:YMEins} vary as
\begin{subequations}
\begin{align} \delta\left(g^{\mu\nu}g^{\alpha\beta}{F_{\mu\alpha}}^a{F_{\nu\beta}^a}\right)\to&0\;,\\
    \delta\left(\frac{1}{2\xi}g^{\mu\nu}g^{\alpha\beta}\mathcal{D}_\alpha A^a_\beta \mathcal{D}_\mu A^a_\nu\right)\to &-\frac{\theta}{g\xi}\tilde{g}^{\mu\nu}\tilde{g}^{\alpha\beta}(\tilde\nabla_\mu A^a_\nu)(\tilde\nabla_\alpha D_\beta c^a)\;,\\
    \delta\left(-\frac{\tMpl}{F(\chi)}\bar{c}^a\tilde{g}^{\mu\nu}\tilde\nabla_\mu D_\nu{c}^a\right)\to &\frac{\theta}{g\xi}\tilde{g}^{\mu\nu}\tilde{g}^{\alpha\beta}(\tilde\nabla_\mu A^a_\nu)(\tilde\nabla_\alpha D_\beta c^a)\;,
\end{align}
\end{subequations}
and the invariance of the Einstein-frame action~\eqref{eq:YMEins} can be verified readily.

    
    \section{Diagonalizing graviton-scalar kinetic mixing}
    \label{appendix_C}
    
    A convenient way to eliminate all the kinetic mixings is to find the matrix transformation that diagonalizes the kinetic terms. However, creating a kinetic mixing matrix between 2-forms (the graviton) and scalar fields is not straightforward. In this appendix, we describe a method for determining the transformation and diagonalizing the kinetic terms, which is implemented in \code{FeynMG} in the function \mmaInlineCell{Code}{\mmaDef{GravKinMixing[]}}.
    
    The main obstacle is that the graviton kinetic term contains both $h_{\mu\nu}$ and its trace $h$. For example, we might have a Lagrangian of the form
    \begin{equation}\label{eq:Lag_kinmixing_ex}
        \lgr=\frac{1}{2}\partial_\rho h_{\mu\nu}\partial^\rho h^{\mu\nu} -\frac{1}{4}\partial_\rho h\partial^\rho h -C\partial_\rho h\partial^\rho \chi+\frac{1}{2}\partial_\rho \chi\partial^\rho \chi, 
    \end{equation}
    where both the graviton and the scalar field have already been canonically normalized, but there remains a kinetic mixing proportional to $C$ (which for the calculation from Section~\ref{Subsection Calculation Jordan} corresponds to $C=\hat{F}(v_\chi)/4$). Since the graviton has two kinetic terms, it is unclear how to construct a matrix that encapsulates all the kinetic couplings between distinct fields.
    
    We proceed by redefining $h_{\mu\nu}$ so that its kinetic energy contains only one term. To do so, we perform an analytic continuation of the graviton into the complex plane, redefining
    \begin{equation}
        \label{eq:complexification}
        h_{\mu\nu}\to\tilde{h}_{\mu\nu}-\frac{1}{4}(1+i)\tilde{h}\eta_{\mu\nu}.
    \end{equation}
    This leads to the Lagrangian
    \begin{equation}
        \lgr=\frac{1}{2}\partial_\rho \tilde{h}_{\mu\nu}\partial^\rho \tilde{h}^{\mu\nu}+ C i\partial_\rho \tilde{h}\partial^\rho \chi+\frac{1}{2}\partial_\rho \chi\partial^\rho \chi,
    \end{equation}
    which contains only one kinetic energy term for the graviton. The kinetic matrix is then defined straightforwardly as
    \begin{equation}\label{eq_KAppendix}
        	K=\begin{pmatrix}
				\frac{1}{2}&  &i\frac{C}{2}\\\\
				i\frac{C}{2}&  &\frac{1}{8}\\
			\end{pmatrix},
    \end{equation}
    with partial derivatives of the fields collected into the vector
    \begin{equation}
        	F_{\rho \mu\nu}=\begin{pmatrix}
				\vspace{-1mm}\partial_\rho\tilde{h}_{\mu\nu} \\  \\\vspace{1mm}\eta_{\mu\nu}\partial_\rho\chi
			\end{pmatrix},
    \end{equation}
    such that the Lagrangian~\eqref{eq:Lag_kinmixing_ex} can be written in the form $\lgr=(F^{\rho\mu\nu})^{\mathsf{T}} K F_{\rho\mu\nu}$, where $\mathsf{T}$ denotes matrix transposition.
    
    We want a transformation $W$ of the matrix $K$ such that 
    \begin{equation}
        W^{\mathsf{T}} K W=\begin{pmatrix}
				\frac{1}{2}&  &0\\\\
				0&  &\frac{1}{8}\\
			\end{pmatrix}.
    \end{equation}
    The transformations for the fields are as follows:
     \begin{equation}
        (F^{\rho\mu\nu})^{\mathsf{T}} K F_{\rho\mu\nu}=(F^{\rho\mu\nu}W^{-1}W)^{\mathsf{T}} K W W^{-1}F_{\rho\mu\nu}=(\tilde{F}^{\rho \mu\nu})^{\mathsf{T}} W^{\mathsf{T}} K W \tilde{F}_{\rho\mu\nu},
    \end{equation}
    since, by defining $\tilde{F}_{\rho \mu\nu}=W^{-1}F_{\rho \mu\nu}$, we would get a Lagrangian free of kinetic mixings.

    For the generic kinetic mixing, where $K$ is defined by Eq.~\eqref{eq_KAppendix}, the transformation matrix is
    \begin{equation}
        W=\begin{pmatrix}
				1&  &\frac{-1}{\sqrt{1+4C^2}}\\\\
				0&  &\frac{-iC}{\sqrt{1+4C^2}}\\
			\end{pmatrix}.
    \end{equation}
    The scalar fields transform through $F_{\rho \mu\nu}=W\tilde{F}_{\rho \mu\nu}$ and therefore
    \begin{subequations}
        \begin{align}
            \tilde{h}_{\mu\nu}&\to\tilde{h}_{\mu\nu} - \frac{iC}{\sqrt{1+4C^2}}\sigma \eta_{\mu\nu},\\
            \chi&\to\frac{-1}{\sqrt{1+4C^2}}\sigma.
        \end{align}
    \end{subequations}
    Undoing the complexification in Eq.~\eqref{eq:complexification}, we obtain the transformations of the original fields that diagonalize the kinetic terms:
    \begin{subequations}
        \begin{align}
            {h}_{\mu\nu}&\to{h}_{\mu\nu} + \frac{C}{\sqrt{1+4C^2}}\sigma \eta_{\mu\nu},\\
            \chi&\to\frac{-1}{\sqrt{1+4C^2}}\sigma.
        \end{align}
    \end{subequations}
    For the specific case of the Lagrangian in \code{Out[16]} from Section~\ref{Example_Jordan}, for which $C=1/\sqrt{2(2\omega+1)}$, we get
    \begin{subequations}
        \begin{align}
            {h}_{\mu\nu}&\to{h}_{\mu\nu} + \frac{1}{\sqrt{2(2\omega+3)}}\sigma \eta_{\mu\nu},\\
            \chi&\to-\frac{\sqrt{2\omega+1}}{\sqrt{2\omega+3}}\sigma.
        \end{align}
    \end{subequations}
    

    \section{Functions of \code{FeynMG}}
    \label{appendix:functions}
    
    
\subsection{Curvature building blocks }
\label{appendix:functions:curvature}

\noindent
\textbf{\code{gUp[i1,i2]}}\hspace{1 em}---\hspace{1 em} Spacetime metric with raised indices, which must be specified as Lorentz, i.e., \code{gUp[Index[Lorentz, i1],Index[Lorentz, i2]]} leads to an upper-indexed metric $g^{i1 i2}$. For more information on the function \code{Index}, see \code{FeynRules} manual~\cite{feynrules}.\\\\
\noindent
\textbf{\code{gDown[i1,i2]}}\hspace{1 em}---\hspace{1 em} Spacetime metric with lowered indices, which must be specified as Lorentz, i.e., \code{gDown[Index[Lorentz, i1],Index[Lorentz, i2]]} leads to an upper-indexed metric $g_{i1 i2}$.\\\\
 \textbf{\code{eta[i1, i2]}}\hspace{1 em}---\hspace{1 em} Flat spacetime metric $\eta^{i1 i2}$. The indices must be Lorentzian, such that \code{eta[Index[Lorentz,i1],Index[Lorentz,i2]]}. The specification of lower or upper indices is not necessary in this case, since \code{FeynRules} does not make that distinction.\\\\
 \textbf{\code{Sqrtg}}\hspace{1 em}---\hspace{1 em} Square root of minus the determinant of the metric, corresponding to the Jacobian factor $\sqrt{-g}$ of the volume element. By default it is assumed as a prefactor to any Lagrangian.\\ \\
 \textbf{\code{VUp[mu,a]}}\hspace{1 em}---\hspace{1 em} Upper indexed vierbein $e^{\mu a}$. Indices must be Lorentzian, such that \code{VUp[Index[Lorentz,mu],Index[Lorentz,a]]}.\\\\
  \textbf{\code{VDown[mu,a]}}\hspace{1 em}---\hspace{1 em} Lower indexed vierbein $e_{\mu a}$. Indices must be Lorentzian, such that \code{VDown[Index[Lorentz,mu],Index[Lorentz,a]]}.\\\\
  \textbf{\code{CovDev[A,mu]}}\hspace{1 em}---\hspace{1 em} Gravitational covariant derivative. As in General Relativity, it will take a different form depending on which object it is acting on (i.e., a spinor, scalar or vector field). \\\\
  \textbf{\code{ChrisSym[a,b,c]}}\hspace{1 em}---\hspace{1 em} Christoffel symbols $\Gamma^{a}_{b c}$ of General Relativity.\\\\
     \textbf{\code{RiemannTensor[a,b,c,d]}}\hspace{1 em}---\hspace{1 em} Riemann curvature tensor (4-form). It will appear in the Lagrangian as $R^{a}_{b c d}$ until the function \code{ExpandGravity} is applied.\\\\
  \textbf{\code{RicciTensor[a,b]}}\hspace{1 em}---\hspace{1 em} Ricci tensor (2-form). It will appear as $R_{a b}$ in the Lagrangian until the function \code{ExpandGravity} is applied.\\\\
  \textbf{\code{RScalar}}\hspace{1 em}---\hspace{1 em} Ricci scalar. It will appear in the Lagrangian as $R_{\rm Sc}$ until the function \code{ExpandGravity} is applied.\\\\
  \textbf{\code{SHGauge[F]}}\hspace{1 em}---\hspace{1 em} Generalization of the harmonic gauge --- the {\it scalar-harmonic gauge}~\cite{Copeland:2021qby}, see Eq.~\eqref{GFterm} --- for Brans--Dicke theories with a curvature term of the form $\lgr_{\rm G}=-\sqrt{-g} \code{F} R/2$. It reduces to the usual harmonic gauge for the Einstein--Hilbert action with $\code{F}=\Mpl^2$.\\\\
  \textbf{\code{CheckMetric[L]}}\hspace{1 em}---\hspace{1 em} Tests whether every pair of indices in a Lagrangian \code{L} is contracted with a metric.\\\\
  \textbf{\code{InsertMetric[L]}}\hspace{1 em}---\hspace{1 em} Takes a Lagrangian \code{L} and inserts an upper indexed metric $g^{\mu \nu}$ at every pair of contracted indices. Useful for adapting \code{FeynRules} model files for use in \code{FeynMG}.\\\\
  \textbf{\code{InsertDevs[L]}}\hspace{1 em}---\hspace{1 em} Upgrades all partial derivatives of vector and fermion fields to covariant derivatives. Useful for adapting \code{FeynRules} model files for use in \code{FeynMG}.\\\\
  \textbf{\code{InsertCurv[L]}}\hspace{1 em}---\hspace{1 em} Applies both \code{InsertMetric} and \code{InsertCurv} to the Lagrangian \code{L}. Useful for adapting \code{FeynRules} Model Files for use in \code{FeynMG}.\\
  

\subsection{Metric transformations}
\label{appendix:functions:metrictransf}

\noindent
  \textbf{\code{ToEinsteinFrame[L,Opts]}}\hspace{1 em}---\hspace{1 em} Performs a Weyl transformation of the Lagrangian \code{L} to the Einstein frame, in the case of Brans--Dicke-type scalar-tensor theories. By specifying the options (\code{Opts}), the user can turn off the default inclusion of the Jacobian $\sqrt{-g}$, using \{\code{Jacobian}$\to$\code{Off}\}.\\\\
  \textbf{\code{WeylTransform[L,$w$]}}\hspace{1 em}---\hspace{1 em} Performs a Weyl transformation of a Lagrangian \code{L}, such that the metric transforms as $g_{\mu\nu}\to w^2 g_{\mu\nu}$ and $g^{\mu\nu}\to w^{-2} g^{\mu\nu}$ and the vierbeins as $e_{\mu}^a\to w e_{\mu}^a$ and $e^{\mu}_a\to w^{-1} e^{\mu}_a$.\\\\
  \textbf{\code{GravityOff[L]}} Takes the Minkowski limit for all curvature objects and eliminates all gravitational perturbations (the graviton) in a Lagrangian \code{L}.\\


\subsection{Expansion tools}
\label{appendix:functions:perturb}

  \noindent \textbf{\code{LinearizeGravity[L,Opts]}}\hspace{1 em}---\hspace{1 em} Linearizes gravity around a flat background metric up to second order in the gravitational sector and first order in the matter sector of a Lagrangian \code{L}. By specifying the options (Opts), the gravitational sector can be linearized up to third order using \{\code{Grav3pt$\to$On}\} and the matter sector up to second order using \{\code{Matter2nd$\to$On}\}. The user can turn off the default inclusion of the Jacobian $\sqrt{-g}$, using \{\code{Jacobian}$\to$\code{Off}\}. Moreover, for Brans--Dicke gravitational sectors, one can choose to automatically use the scalar-harmonic gauge (\code{SHGauge}) using \{\code{SHGauge$\to$ On}\}, and update the rest of the covariant derivatives into their modified form from Eq.~\eqref{eq_ModifiedCovDev} using \{\code{UpdDevs$\to$ On}\}.\\\\
  \textbf{\code{ExpandGravity[L]}}\hspace{1 em}---\hspace{1 em} Expands all the gravitational objects, such as the Ricci scalar, Ricci tensor or Riemann tensor in terms of the metric.\\\\
  \textbf{\code{ExpandCMod[L]}}\hspace{1 em}---\hspace{1 em} Expands the \code{CMod} (modification of the covariant derivatives) in terms of the scalar degree of freedom. This function will be automatically applied once all the scalar fields are canonically normalized.\\\\
  \textbf{\code{Orderh[L,n]}}\hspace{1 em}---\hspace{1 em} Truncates a Lagrangian \code{L} up to the $n$-th order in the perturbation of the metric perturbation $h_{\mu\nu}$.\\\\
  \textbf{\code{OrderSimplify[L,n]}}\hspace{1 em}---\hspace{1 em} Applies the \code{Mathematica} function \code{FullSimplify} to all the terms in a Lagrangian \code{L} of $n$-th order or lower in the fields.\\\\
  \textbf{\code{IndexSimplify[L,\{i1,i2,...\}]}}\hspace{1 em}---\hspace{1 em} Replaces the Lorentz indices of all the terms of a Lagrangian \code{L} so that equivalent terms can be combined. The second argument allows the user to specify a set of indices from which the replacements will be chosen.\\\\
  \textbf{\code{IndexChange[L,\{i1,i2,...\}]}}\hspace{1 em}---\hspace{1 em} Replaces the Lorentz indices of a Lagrangian \code{L} sequentially from the set of indices \code{\{i1,i2,...\}}.\\


\subsection{Tools for canonicalizing fields}
\label{appendix:functions:canon}

\noindent
  \textbf{\code{CanonScalar[L]}}\hspace{1 em}---\hspace{1 em} Canonically normalizes the leading kinetic energy terms of the scalar fields of the Lagrangian \code{L}.\\\\
  \textbf{\code{CanonGrav[L]}}\hspace{1 em}---\hspace{1 em} Canonically normalizes the graviton kinetic energy, assuming that the kinetic terms have a constant prefactor.\\\\
  \textbf{\code{MassDiagMG[L]}}\hspace{1 em}---\hspace{1 em} Diagonalizes the scalar field mass matrix of the Lagrangian \code{L}.\\\\
  \textbf{\code{KineticDiagMG[L,n]}}\hspace{1 em}---\hspace{1 em} Diagonalizes the kinetic energies of the scalar fields of the Lagrangian \code{L}.\\\\
  \textbf{\code{GravKinMixing[L]}}\hspace{1 em}---\hspace{1 em} Diagonalizes the kinetic mixings between the trace of the graviton and the scalar fields of the Lagrangian \code{L}.\\


\subsection{Vacuum expectation values}
\label{appendix:functions:vevs}

\noindent
  \textbf{\code{VevExtract[L,Opts]}}\hspace{1 em}---\hspace{1 em} Solves for the vacuum expectation values of the real scalar fields in the Lagrangian \code{L}. By specifying the options \{\code{Fields$\to$} \code{\{p1,p2,...\}}\}, the user can choose which fields to expand around their vevs.\\\\
  \textbf{\code{VevExpand[L]}}\hspace{1 em}---\hspace{1 em} Expands and solves for the vacuum expectation values of the real scalar fields in the Lagrangian \code{L}. The function will output all the different solutions, and a dialogue window will prompt the user to select a set of vevs for substitution into the Lagrangian. By specifying the options \{\code{Solution$\to$n}\} and \{\code{Fields$\to$\{p1,p2,...\}}\}, the user can choose the $n$-th solution and the fields to be expanded directly.\\\\


\subsection{Checking functions}
\label{appendix:functions:check}

\noindent
  \textbf{\code{CheckCanonScalar[L]}}\hspace{1 em}---\hspace{1 em} Finds the leading scalar field kinetic energy terms in the Lagrangian \code{L} and tests whether they are canonically normalized.\\\\
  \textbf{\code{CheckMassMatrix[L]}}\hspace{1 em}---\hspace{1 em}  Extracts the mass matrix for the scalar fields of the Lagrangian \code{L} and checks if it is diagonalized.\\\\
  \textbf{\code{CheckKineticMatrix[L]}}\hspace{1 em}---\hspace{1 em}  Extracts the kinetic energy matrix for the scalar fields of the Lagrangian \code{L} and checks if it is diagonalized.\\\\
  \textbf{\code{CheckGravityMixing[L]}}\hspace{1 em}---\hspace{1 em}  Checks whether there is any kinetic mixing between the trace of the graviton $h$ and a scalar field.\\


\subsection{Effective Planck mass}
\label{appendix:functions:mpl}

\noindent
  \textbf{\code{GiveMpl[L]}}\hspace{1 em}---\hspace{1 em} Extracts the effective $\Mpl$ from the Lagrangian \code{L}. It can be used at any stage of the calculation (before or after linearizing gravity or canonically normalizing the kinetic energies).\\\\
 \textbf{\code{InsertMpl[L]}}\hspace{1 em}---\hspace{1 em} Extracts and inserts the effective $\Mpl$ of the Lagrangian \code{L}. It can be used at any stage of the calculation (before or after linearizing gravity or canonically normalizing the kinetic energies).\\\\


\subsection{Output model file}
\label{appendix:functions:output}

\noindent
  \textbf{\code{AddScalar[P,Opts]}}\hspace{1 em}---\hspace{1 em} Adds a new massless scalar field named \code{P} into the loaded set of fields, such that it can be recognized by \code{FeynRules}. Within the options (\code{Opts}), the user can choose the mass and width of the associated particle by including \code{\{Mass$\to$X\}} or \code{\{Width$\to$X\}}, respectively.\\\\
  \textbf{\code{AddParameter[P,Opts]}}\hspace{1 em}---\hspace{1 em} Adds a new parameter named \code{P} into the loaded set of parameters, such that it can be recognized by \code{FeynRules}. Within the options (\code{Opts}), the user can choose its value by including \code{\{Value$\to$X\}} or its interaction order by including \code{\{InteractionOrder$\to$X\}} (where $X$ is set to 1 by default).\\\\
  \textbf{\code{OutputModelMG[OldF,NewF,L,Opts]}}\hspace{1 em}---\hspace{1 em} Creates a new model file named \code{NewF} from an original \code{FeynRules} model file \code{OldF}. The new model file will contain the same defined fields and parameters as the original file, with the addition of all the new particles and parameters created using \code{AddScalar} and \code{AddParameter}, together with the Lagrangian (\code{L}), the graviton ($h_{\mu\nu}$) and Planck mass ($\Mpl$). By specifying the option  \code{\{UpdateMass$\to$True\}}, the masses of all scalar fields will be updated.

\end{appendix}
	

\end{document}